\documentclass[11pt,titlepage]{article}
\usepackage{graphicx}
\usepackage{amssymb}
\usepackage{amsmath}

\renewcommand{\theequation}{\thesection.\arabic{equation}}

\font\medio=cmr10 scaled \magstep2
\outer\def\beginsection#1\par{\medbreak\bigskip
      \message{#1}\leftline{\bf#1}\nobreak\medskip
\vskip-\parskip
      \noindent}

\usepackage{color}
\def\laq{\raise 0.4ex\hbox{$<$}\kern -0.8em\lower 0.62
ex\hbox{$\sim$}}
\def\gaq{\raise 0.4ex\hbox{$>$}\kern -0.7em\lower 0.62
ex\hbox{$\sim$}}
\def\beq{\begin{equation}}
\def\eeq{\end{equation}}
\def\bea{\begin{eqnarray}}
\def\eea{\end{eqnarray}}
\def\bean{\begin{eqnarray*}}
\def\eean{\end{eqnarray*}}

\def \be {\begin{equation}}
\def \ee {\end{equation}}
\def \Hcal {{\cal H}}

\def \pa {\partial}
\def \ra {\rightarrow}

\def \la {\lambda}

\def \ga {\gamma}

\def \da {\delta}
\def \ep {\epsilon}

\def \Om {\Omega}

\def \pa {\partial}

\def \ra {\rightarrow}

\def \la {\lambda}

\def \ga {\gamma}

\def \da {\delta}
\def \ep {\epsilon}

\def \Om {\Omega}

\begin{document}
\bibliographystyle {unsrt}

\titlepage

\begin{flushright}
CERN-PH-TH/2013-196 \\
BA-TH/675-13\\

\end{flushright}
\vspace{8mm}
\begin{center}
{\huge \bf  An exact Jacobi map}\\
\vspace{0.4 cm}
{\huge \bf   in the  geodesic light-cone gauge}\\

\vspace{10mm}

\vspace{3mm}
{  G. Fanizza$^{1,2}$, M. Gasperini$^{1,2}$, G. Marozzi$^{3}$ and G. Veneziano$^{4,5}$} \\
\vspace{6mm}
{\sl ${ }^{1}$Dipartimento di Fisica, Universit\`a di Bari, Via G. Amendola 173\\70126 Bari, Italy}\\
\vspace{3mm}
{\sl ${ }^2$Istituto Nazionale di Fisica Nucleare, Sezione di Bari, Bari, Italy}\\
\vspace{3mm}
{\sl ${ }^{3}$Universit\'e de G\`eneve, D\'epartement de Physique Th\'eorique and CAP,\\
24 quai Ernest-Ansermet, CH-1211 Gen\`eve 4, Switzerland}\\
\vspace{3mm}
{\sl ${ }^{4}$Coll\`ege de France, 11 place Marcelin  Berthelot \\
 75005 Paris, France}\\ 
\vspace{3mm}
{\sl ${ }^{5}$Theory Division, CERN, CH-1211 Geneva 23, Switzerland}

\end{center}

\vskip 1.5cm
\centerline{\medio  Abstract}

The remarkable properties of the recently proposed geodesic light-cone  (GLC) gauge allow to explicitly solve the geodesic-deviation equation, and thus to derive  an exact  expression for the Jacobi map $J^A_B(s,o)$ connecting a generic source $s$ to a geodesic observer $o$ in a generic space time. In this gauge $J^A_B$  factorizes into the product of a local quantity at $s$ times one at $o$, implying  similarly factorized expressions for the area and luminosity distance. In any other coordinate system $J^A_B$  is simply given by expressing the GLC quantities in terms of the corresponding ones in the new coordinates. This is explicitly done, at first and second order, respectively, for the synchronous and Poisson gauge-fixing of a perturbed, spatially-flat cosmological background, and the consistency of the two outcomes is  checked.
Our results slightly amend previous calculations of the luminosity-redshift relation and suggest a possible non-perturbative way for computing the effects of inhomogeneities on observations based  on light-like signals.

\newpage

%%%%%%%%%%%%%%%%%%%%%%%%%%%%%%%%%%%%%%%%%%%%%%%%%%%%%%%%%%
\section{Introduction}
\label{Sec1}
\setcounter{equation}{0}
%%%%%%%%%%%%%%%%%%%%%%%%%%%%%%%%%%%%%%%%%%%%%%%%%%%%%%%%%%

Most cosmological observations (cosmic rays and massive neutrinos being worthy exceptions) are based on light-like signals received by an observer  moving on his/her own world-line. The signals thus travel on the past light cones with tips on the aforementioned worldline. It is clear that a coordinate system adapted to the observer {\it and} to his/her past light cones can greatly simplify the computation of the  expected signals within a specific theoretical framework.
The geodesic light-cone  (GLC) gauge was indeed introduced in \cite{GMNV} in order to cope in a most efficient way with this class of problems, and was later applied, in particular, to the study of the luminosity/redshift relation in the presence of inhomogeneities \cite{BGMNV1, BGMNV2, BMNV, BGMNV3}. 

 Although in \cite{GMNV} it was superficially stated that the GLC gauge is a complete gauge fixing of the so-called observational coordinates  defined in \cite{Maartens1,Maartens3}   (see also \cite{Clarkson}), this is actually incorrect. Indeed, the GLC coordinates consist of a timelike  coordinate $\tau$ (which can always be  identified with the proper time of the synchronous gauge \cite{BGMNV1}),  of a null coordinate $w$ and of two angular coordinates $\tilde{\theta}^a$ ($a=1,2$). The timelike coordinate $\tau$ thus replaces the spacelike coordinate $y$ of \cite{Maartens1,Maartens3}, leading to important qualitative  differences.

The line-element of the GLC metric takes the form:  
\beq
\label{LCmetric}
ds^2 =\Upsilon^2 dw^2 - 2 \Upsilon  dw d\tau+\gamma_{ab}(d \tilde{\theta}^a-U^a dw)(d \tilde{\theta}^b-U^b dw) ~~,~~~~~ a, b = 1,2~~,
 \eeq
 and thus depends on six arbitrary functions ($\Upsilon$,  $U^a$ and $\gamma_{ab} = \gamma_{ba}$).
In matrix form:
\beq
\label{GLCmetric}
g_{\mu\nu} =
\left(
\begin{array}{ccc}
0 & -\Upsilon &  \vec{0} \\
-\Upsilon & \Upsilon^2 + U^2 & -U_b \\
\vec0^{\,T}  &-U_a^T  & \gamma_{ab} \\
\end{array}
\right)
~~~~~,~~~~~
g^{\mu\nu} =
\left(
\begin{array}{ccc}
-1 & -\Upsilon^{-1} & -U^b/\Upsilon \\
-\Upsilon^{-1} & 0 & \vec{0} \\
-(U^a)^T/ \Upsilon & \vec{0}^{\, T} & \gamma^{ab}
\end{array}
\right) ~,
\eeq
where $\gamma_{ab}$ and its inverse  $\gamma^{ab}$ lower and  raise  the two-dimensional indices. 

In analogy with the synchronous gauge also the GLC gauge has some residual gauge freedom. An interesting example, that we shall be using later, consists of
the  (finite) coordinate transformation:
\be
\label{GLCres}
\tau \ra \tau ~~~;~~~ w \ra w~~~;~~~ \tilde{\theta}^a \ra \bar{\theta}^a( \tilde{\theta}^b, w)\, .
\ee
This preserves  the GLC gauge, changing both $\gamma_{ab}$ and $U^a$, and leaving $\Upsilon$ unchanged.

 The condition $w=$ constant defines a null hypersurface (since  $\pa_\mu w \pa^\mu w=0$), corresponding to the past light-cone of a particular observer (in practice chosen to be ourselves). The vector $u_\mu = - \partial_{\mu} \tau$ describes a geodesic flow, since $\left( \pa^\nu \tau\right) \nabla_\nu \left( \pa_\mu \tau\right) = 0$, 
and, as mentioned, is associated to geodesic observers which are static in the synchronous gauge \cite{BGMNV1} . Let us also recall  that, in the GLC gauge, the null geodesics connecting sources and observer are characterized by the simple tangent vector $k^{\mu} = - \omega g^{\mu \nu} \partial_{\nu} w = -  \omega g^{\mu w} = \omega  \Upsilon^{-1} \delta^{\mu}_{\tau}$ (where $ \omega$ is an arbitrary normalization constant), meaning that photons  travel at constant values of $w$ and $\tilde{\theta}^a$. This makes this gauge particularly adapted to the computation of the redshift $z$ and of the area distance $d_A$  \cite{GMNV, BMNV}, and also simplifies  the task of computing the light-cone 
average\footnote{Introduced in \cite{GMNV} by extending to null hypersurfaces the gauge invariant procedure for space-like domains defined 
in \cite{GMV1,GMV2}.}
and dispersion of various  observables connected to the luminosity-redshift relation \cite{BGMNV1, BGMNV2,  BGMNV3}.

In this paper we shall extend those results to the construction of the fully non-perturbative Jacobi map (JM) linking a generic source to a geodesic  observer in a generic inhomogeneous and anisotropic cosmological setup. As we shall see, this more rigorous treatment slightly amends the starting expression for the luminosity distance used in \cite{BGMNV1, BMNV}:  the dependence of the result on the source is exactly the same, while the dependence on the variables at the observer needs a little qualification. 
 In essence, depending on the gauge considered, there is an additional contribution that can be interpreted in terms of the relativistic change 
of the solid angle due to the observer's peculiar velocity and/or to some anisotropy at the observer as measured in that gauge.
 Namely, in terms of its dependence on the above quantities, $J^A_B$ will {\it look} different in different gauges,  even if the full expression is actually the same.

The paper is organized as follows. In Sect. \ref{Sec2} we recall the definition of the JM and the basic coordinate-independent equations it should satisfy. In Sect. \ref{Sec3} we derive our main result: an exact expression for the JM and for the area/luminosity distance in the GLC gauge. We find that the expression used in previous work emerges after using the residual gauge symmetry in Eq. (\ref{GLCres}). In Sect. \ref{Sec4}, considering the case of a static geodesic observer, we shall express the area distance in terms of  synchronous gauge (SG) quantities, by using its first-order relation with the GLC gauge. No relativistic (Doppler) correction at the observer emerges (consistently with the fact that the observer is static in the SG); we find, rather, an ``anisotropy correction" that can be ``gauged-away" by an appropriate redefinition of the coordinates  around the observer, which is allowed by the residual gauge freedom of the SG. In Sect. \ref{Sec5} we use the  first-order coordinate transformation between the Poisson gauge (PG) and the GLC gauge in order to write the area distance in terms of PG quantities: we notice the emergence of new terms (calculated also to second-order) 
that we can interpret as relativistic corrections due to the peculiar velocity of the geodesic observer in the PG.  
In Sect. \ref{Sec6} we compare the SG and PG results at first order and show that there is full agreement between them provided, once more, the residual gauge freedom of the SG is appropriately fixed.  In Sect. \ref{Sec7} we summarize our main results and draw a few conclusions.
 In Appendix A we give, for convenience, some useful formulae pertaining to the GLC gauge and  discuss how one can impose a property of the two-dimensional Sachs basis used in this paper. Finally, in Appendix B, we give, for completeness (and also in view of  possible phenomenological applications), the (slightly amended with respect to \cite{BMNV}) explicit second-order expression for the  luminosity distance as a function of the observed redshift in the PG,
showing how to recover agreement with the first-order results available in the literature. 

%%%%%%%%%%%%%%%%%%%%%%%%%%%%%%%%%%%%%%%%%%%%%%%%%%%%%%%%%%%%%%%%%%%%%%%%%%%%%%%%%%%%%%%%%%%%%%%%%%%%%%%%%%%%%%%%%%%
%%%%%%%%%%%%%%%%%%%%%%%%%%%%%%%%%%%%%%%%%%%%%%%%%%%%%%%%%%%%%%%%%%%%%%%%%%%%%%%%%%%%%%%%%%%%%%%%%%%%%%%%%%%%%%%%%%%

%%%%%%%%%%%%%%%%%%%%%%%%%%%%%%%%%%%%%%%%%%%%%%%%%%%%%%%%%%%%%%%%%%%%%%%%%%%%%%%%%%%%%%%%%%%%%%%%%%%%%%%%
\section{Definition of the Jacobi Map}
\label{Sec2}
\setcounter{equation}{0}
%%%%%%%%%%%%%%%%%%%%%%%%%%%%%%%%%%%%%%%%%%%%%%%%%%%%%%%%%%%%%%%%%%%%%%%%%%%%%%%%%%%%%%%%%%%%%%%%%%%%%%%%%%

In the following we shall basically use the notations and conventions of \cite{FDU} (see also \cite{CL,Bonvin}). Let us consider the geodesic deviation equation:
\begin{equation}
\label{eq:geodesicDeviation}
\nabla_{\la}^2 \xi^\mu= {R_{\alpha\beta\nu}}^\mu k^\alpha k^\nu \xi^\beta\, ,
\end{equation}
where ${\la}$ is the affine parameter along the geodesics, and $\nabla_{\la} \equiv  k^\alpha\nabla_\alpha$.
This equation concerns displacements $\xi^{\mu}$ which are orthogonal to  $k^{\mu}$. However, since the equation for 
the component of $\xi^{\mu}$  along $k^{\mu}$ (which obeys the orthogonality conditions as $k^{\mu}$ is a null vector) is trivially satisfied, we are left with only two  components of $\xi^{\mu}$  satisfying non-trivial equations.
Therefore, without any lack of generality, we can project $\xi^\mu$  along the so-called Sachs basis $\{s_A^\mu\}$ \cite{Sachs2,SSE}, 
namely along two parallely transported 4-vectors  $s_A^\mu$ ($A = 1,2$)  defined by the conditions \cite{FDU,PUP}:
\begin{align}
\label{eq:SachsBasis1}
g_{\mu\nu}s_A^\mu s_B^\nu&=\delta_{AB}  \,,\\
\label{eq:SachsBasis2}
s_A^\mu u_\mu =0 ~~~~~~~~~~~~~~&,~~~~~~~~~~~~~~
s_A^\mu k_\mu =0,\\
\label{eq:SachsBasis4}
\Pi^\mu_\nu\nabla_{\la} s_A^\nu=0 ~~~~~  \text{with} ~~~~~ \Pi^\mu_\nu &= \delta^\mu_\nu - \frac{k^\mu k_\nu}{(u^\alpha k_\alpha)^2} - \frac{k^\mu u_\nu + u^\mu k_\nu}{u^\alpha k_\alpha} \, ,
\end{align}
where $\Pi^\mu_\nu$ is a projector on the two-dimensional space orthogonal to $u_\mu$ and to $n_\mu = u_\mu + (u^\alpha k_\alpha)^{-1} k_\mu$ with $n^\alpha n_\alpha = 1$ and $n^\alpha k_\alpha = 0$.
In other words, neglecting terms proportional to $k^{\mu}$, we can write $\xi^\mu$ in the form
\beq
\xi^\mu =  \xi^A s_A^{\mu} \, ,
\label{ximusmu}
\eeq
with  (the height of the flat $A,B, \dots$ indices is irrelevant)
\beq
\xi^A = \xi^\mu s^A_{\mu} =  g_{\mu\nu} \xi^\mu s_A^{\nu} \,.
\eeq
 We thus obtain\footnote{Note that derivatives now act on a spacetime scalar.}:
\begin{equation}
\label{eq:geodesicDeviationOnSachs}
\frac{d^2\xi^A}{d{\la}^2}  =   R^A_B \xi^B ~~~~~~~~~~~~  ,
~~~~~~~~~~~~ \frac{d}{d{\la}} \equiv k^{\mu} \pa_{\mu}\,,
\end{equation}
where we have defined
\begin{equation}
\label{eq:RiemannOnSachs}
R^A_B \equiv R_{\alpha \beta \nu\mu} k^\alpha k^\nu s_B^\beta s_A^{\mu} \, ,
\end{equation}
in agreement with eq. (3.7) of \cite{FDU}.

The Jacobi map  \cite{SEF} $J^A_B(\la_s,\la_o)$, connecting an observer $o$ to a source $s$, is the solution of Eq. (\ref{eq:geodesicDeviationOnSachs}) expressed in the form:
\beq
\label{defJ}
\xi^A(\la_s) =  J^A_B({\la_s},{\la}_o) \left( \frac{k^{\mu} \pa_{\mu}\xi^B}{k^{\nu} u_{\nu}} \right)_o \, 
\eeq
(hereafter we denote by a suffix $o$ quantities defined at the observer position, identified with the origin of the geodesic bundle, and by a suffix $s$ quantities defined at the source position). 
Here  $u_{\mu}$ is the observer's 4-velocity, and the 2x2 matrix  $J^A_B$ satisfies:
\beq
\label{eq:JMequation}
\frac{d^2 }{d\la^2} J^A_B(\la, \la_o) = R^A_C  J^C_B \, ,
\eeq
with the initial conditions
\beq
\label{eq:JMinitialcondition}
 J^A_B(\la_o,\la_o) = 0 ~~;~~~~~~~ \frac{d }{d\la} J^A_B(\la_o,\la_o) = \delta^A_B \, (k^\nu u_\nu)_o~.
\eeq
Knowledge of $ J^A_B$ allows to compute,  among other things, the area distance $d_A$ defined by \cite{Sachs}
\be 
d_A^2\equiv \frac{d S_s}{d \Omega_o},
\label{defdA2book}
\ee
where $d\Omega_o$ is the infinitesimal solid angle subtending the source at the observer position, and $dS$ is the cross-sectional area element perpendicular to the light ray at the source position.
This is connected to the determinant of $J$ by the following well-known relation \cite{SEF}
\beq
\label{JdArel}
d_A^2 = \det J^A_B( \la_s,\la_o).
\eeq

 Quite often (see e.g. \cite{FDU}) one normalizes $k^{\mu}$ by imposing $(k^{\nu} u_{\nu})_o=1$.   We prefer to use the more general form in Eq. (\ref{defJ}) which shows explicitly that $J$ is invariant  under general  transformations of the coordinates at the observer and under a rescaling of $\omega$ (i.e. of $k^{\mu}$) but, unlike Eq. (\ref{eq:geodesicDeviationOnSachs}), depends explicitly upon the observer's velocity. This latter point will be discussed in the next section.
  
In general, solving the Eqs. (\ref{eq:JMequation}) and (\ref{eq:JMinitialcondition}) is a highly non trivial task which can only be done at some low order in perturbation theory, around a particularly symmetric background. Instead, as we shall see below, exact non-perturbative solutions can be obtained by using the GLC gauge, i.e. coordinates that are adapted to observations taking place over the past  light-cone of a geodesic observer.

%%%%%%%%%%%%%%%%%%%%%%%%%%%%%%%%%%%%%%%%%%%%%%%%%%%%%%%%%%%%%%%%%%%%%%%%%%%%%%%%%%%%%%%%%%%%%%%%%%%%%%%%%%%%%%%%%%%
%%%%%%%%%%%%%%%%%%%%%%%%%%%%%%%%%%%%%%%%%%%%%%%%%%%%%%%%%%%%%%%%%%%%%%%%%%%%%%%%%%%%%%%%%%%%%%%%%%%%%%%%%%%%%%%%%%%

%%%%%%%%%%%%%%%%%%%%%%%%%%%%%%%%%%%%%%%%%%%%%%%%%%%%%%%%%%%%%%%%%%%%%%%%%%%%%%%%%%%%%%%%%%%%%%%%%%%%%%%%%%%%%%
\section{An exact Jacobi map in the  GLC gauge}
\label{Sec3}
\setcounter{equation}{0}
%%%%%%%%%%%%%%%%%%%%%%%%%%%%%%%%%%%%%%%%%%%%%%%%%%%%%%%%%%%%%%%%%%%%%%%%%%%%%%%%%%%%%%%%%%%%%%%%%%%%%%%%%%%%%%%

In the GLC gauge  we have $k^\mu \sim  \delta^{\mu}_{\tau}$ and therefore Eq. (\ref{defJ}) becomes, in general:
\beq
\label{defJGLC}
\xi^A(\la_s) =  J^A_B({\la_s},{\la}_o) \left(u_{ \tau}^{-1} \pa_{\tau}\xi^B \right)_{\la_o} \, .
\eeq
 From the condition $s_A^\mu k_\mu=0$ (see Eq. \eqref{eq:SachsBasis2}) we  obtain
\begin{equation}
\label{s^mu}
s_A^\mu=(s_A^{\tau},0,s_A^a),
\end{equation}
where   $a = 1,2 $ refers to the indices of $\gamma_{ab}$.
The remaining conditions in Eqs. (\ref{eq:SachsBasis1}) and (\ref{eq:SachsBasis4}) lead to identify $s^A_a$ with some conveniently chosen zweibeins for  $\gamma_{ab}$. Indeed,  from Eqs. (\ref{eq:SachsBasis1}) and (\ref{s^mu}), and using the form of the GLC metric, we find:
\beq
 g_{\mu \nu}s_A^{\mu} s_B^\nu =  \gamma_{ab} s_A^a s_B^b  =  \delta_{AB}~,
\label{CondZweibeins2} 
\eeq
and therefore the inverse 2x2 matrices  $s^A_a$ satisfy:
\beq
s^A_a s^A_b = \gamma_{ab} \,. 
\label{CondZweibeins1} 
\eeq
Consequently, we also find that
\begin{equation}
\label{s_mu}
s^A_\mu= g_{\mu \nu}s_A^{\nu} = (0, s^A_{w}, s^A_a)\,.
\end{equation}
Finally,  the last condition to be satisfied, Eq. \eqref{eq:SachsBasis4}, can be easily shown to reduce, in the GLC gauge to:
\beq
\label{par}
  k^\nu\nabla_\nu s_A^a =  0 \, .
\eeq
This condition can be easily implemented by using a residual local $U(1)$ rotation of the zweibeins (see Appendix A). 
%On the other hand, $s_A^\tau$ is determined on the observer's worldline by the condition
% $s_A^\mu u_\mu =0$ (as $s_A^\tau = - u_\tau^{-1}s_A^a u_a$) and it is  then parallely transported along the null geodesic through $ k^\nu\nabla_\nu s_A^{\tau} =  0$.
% The actual expression for $s_A^\tau$ will not be needed in the following.

Let us now show that, with the $\xi^a$ constant along the null geodesics, the geodesic deviation equation is automatically satisfied,  and the exact JM can be explicitly given.
Using the results of Appendix A (in particular, ${\Gamma_{\tau \tau}}^\rho=\Upsilon^{-1} \partial_\tau \Upsilon\delta^{\rho}_{ \tau}$ and ${\Gamma_{\tau a}}^w=0$) we can show, first of all, that, for a constant $\xi^a$,  the l.h.s. of Eq. (\ref{eq:geodesicDeviation}) reads 
\beq
\label{nablaxib}
 \nabla_{\la}^2 \xi^b = \left(\frac{\omega}{\Upsilon}\right)^2 \left( \partial_\tau{\Gamma_{\tau a}}^b+{\Gamma_{\tau a}}^c{\Gamma_{\tau c}}^b- {\Gamma_{\tau a}}^b \Upsilon^{-1} \partial_\tau \Upsilon \right) \xi^a.
 \eeq
 On the other hand, an explicit calculation of the only component of the Riemann tensor that contributes to the r.h.s. of 
 Eq. (\ref{eq:geodesicDeviation}) gives:
 \beq
 {R_{\tau a \tau}}^b = \partial_\tau{\Gamma_{\tau a}}^b+{\Gamma_{\tau a}}^c{\Gamma_{\tau c}}^b-
 {\Gamma_{\tau a}}^b \Upsilon^{-1} \partial_\tau \Upsilon \,.
 \eeq
Contracting now this identity with $\xi^a$, and using Eq. (\ref{nablaxib}),  we clearly reproduce Eq. (\ref{eq:geodesicDeviation}) for 
$\mu = b$. 
 In a similar way, we can easily show that the same geodesic deviation equation is satisfied for a constant $\xi^w$ (in our case, from 
Eqs. (\ref{ximusmu}) and (\ref{s^mu}), $\xi^w=0$). 
 
As in the general case (see Section 2), the fact that $s^A_a$ are covariantly constant  also implies  Eq. (\ref{eq:geodesicDeviationOnSachs}) for the $\xi^A$. 
 At this point, using  $\xi^A = \xi^{a} s^A_a + \xi^w s^A_w = \xi^{a} s^A_a$ and  the fact that $\xi^a$ is a arbitrary constant vector, we can also easily prove that:
 \begin{equation}
\label{eq:GeodesicSachsBasis}
\frac{d^2 s_a^A}{d\la^2} =   {R_{\alpha a \beta}}^j k^{\alpha} k^{\beta} s_j^A = {R_{\alpha a \beta}}^b k^{\alpha} k^{\beta} s_b^A \, , ~~~~~~~~~~~ j = (w, a )~,
\end{equation}
where we have taken into account that the term with $j = w$ does not contribute (using again  the affine connections given in Appendix A).
The basic result in Eq. (\ref{eq:GeodesicSachsBasis}) allows us to construct the JM  using the following ansatz:
\beq
\label{eq:JacobiInGLCG}
J^A_B(\la,\la_o) = s^A_a(\la)\,C^a_B, 
\eeq
where $C^a_B$ is a $\lambda$-independent matrix.
Inserting it  in Eq. (\ref{eq:JMequation}) one easily finds that the latter is satisfied as a consequence of Eq. (\ref{eq:GeodesicSachsBasis}).
 Furthermore, the first initial condition in Eq. \eqref{eq:JMinitialcondition} is also satisfied since, by the definition of the origin of the bundle, $s_a^A(\lambda_o)=0$.  We are then left with  fixing the constant $C^a_B$ matrix. To this purpose we  impose the second initial condition of Eq. \eqref{eq:JMinitialcondition}, and  obtain:
\begin{equation}
\left( C^{-1}\right)_a^B = \left(u_{\tau}^{-1} \pa_{\tau} s_a^B\right)_{\la=\la_o} =\left(\frac{k^{\mu}\pa_{\mu}  s_a^B}{k^{\mu} u_{\mu}}
\right)_{\la=\la_o}  \,.
\end{equation}
In conclusion, the final exact expression for the Jacobi map in the GLC gauge can be written in the form:
\beq
 \label{finalJM}
 J^A_B(\la,\la_o) = s^A_a(\la)\left\{\left[\left(\frac{k^{\mu}\pa_{\mu}  s}{k^{\mu} u_{\mu}}\right)^{-1}\right]^a_B\right\}_{\la=\la_o} = s^A_a(\la)\left\{\left[\left(u_{\tau}^{-1}\pa_{\tau}  s\right)^{-1}\right]^a_B\right\}_{\la=\la_o},
 \eeq
 which is the main result of this paper~\footnote{We stress that all the properties of the GLC gauge have been used in our derivation. Whether a similar exact formula for the JM can be derived using the observational coordinate of \cite{Maartens1,Maartens3} (see Sect. 1) remains an interesting and non-trivial open question.}.
 
The above expression clearly includes possible aberration effects due
to the dependence of the solid angle $d\Om_o$ on the observer's peculiar velocity \cite{SEF}.
Indeed, if $u_\mu$ and $\widetilde{u}_\mu$ are the 4-velocities of two different observers, the corresponding solid angles $d\Omega$ and $d\widetilde{\Omega}$ are related by:
\be 
\frac{d\widetilde{\Omega}}{d{\Omega}}=\left(\frac{k_\mu u^\mu}{k_\mu \widetilde{u}^\mu}\right)^2,
\ee
and the results for  $d_A^2$ (see Eq. (\ref{defdA2book})) consequently differ by the factor $ \left({k_\mu u^\mu}/{k_\mu \widetilde{u}^\mu}\right)^{-2}$.
On the other hand, from the result in Eq. (\ref{finalJM}), we have \footnote{The JM depends on $u_\mu$ only through the $k^\mu u_\mu$ factor, as the condition $s^\mu_A u_\mu=0$ only affects the $s_A^\tau$ component of $s_A^\mu$, which does not contribute to the JM.} 
\be 
J^A_B({\la_s},{\la}_o;\tilde{u}_{\mu} ) = \frac{(k^{\mu} \tilde{u}_{\mu})_o}{(k^{\nu} u_{\nu})_o}  J^A_B({\la_s},{\la}_o;u_{\mu} ),
 \eeq
implying that $d_A^2$ changes exactly as prescribed by the aberration effect (see Eq. (\ref{JdArel})). 
 
Let us now give the explicit expression for the area distance $d_A$ through Eq. (\ref{JdArel}):
\beq
\label{areadist}
d_A^2 = \det \left( J^A_B(\la_s,\la_o)\right) = 
 \frac{\sqrt{\gamma(\la_s)}}{\det \left(u_{\tau}^{-1} \partial_\tau s_b^B \right)_{\la= \la_o}} ,~~~~~~~~ \ga \equiv  \det \ga_{ab}\,. 
\eeq
Recalling the relation between the luminosity and area distance through the redshift $z$ \cite{Et1933}, and using the fact that the redshift itself factorizes in the GLC gauge \cite{GMNV}, we also find the following closed expression for $d_L$:
\beq
\label{lumdist}
d_L^2 =  (1+z)^4 d_A^2 = 
 \frac{\Upsilon_s^{-4}\,\sqrt{\gamma(\la_s)}}{\Upsilon_o^{-4}\,\det \left(u_{\tau}^{-1} \partial_\tau s_b^B \right)_{\la= \la_o}} \,. 
\eeq
We can also rewrite the denominators of Eqs. \eqref{areadist} and \eqref{lumdist} as follows:
\begin{equation}
\det \left(u_{\tau}^{-1} \partial_\tau s_b^B \right)_{\la= \la_o} = \frac{1}{4}\left[ \frac{\det \left(u_{\tau}^{-1} \pa_\tau{\gamma}_{ab}\right)}{\sqrt{\gamma}}\right ]_{o}\, ,
\end{equation}
from which we arrive at the useful relation
\begin{equation}
\det \left(u_{\tau}^{-1} \partial_\tau s_b^B \right)_{\la= \la_o}= \frac{1}{4} \left[\det \left(u_{\tau}^{-1} \pa_\tau{\gamma}^{ab}\right) \gamma^{3/2}\right]_{o}\, .
\label{ExpDetS2a}
\end{equation}
From Eqs. (\ref{finalJM}), (\ref{areadist}) and (\ref{lumdist}) we see that, in the GLC gauge, Jacobi map, area and luminosity distance neatly factorize into a local contribution at the source times a local contribution at the observer. Note that their expressions are {\it not} general covariant and, as such, cannot be used in other coordinate systems (where they will take, in general, very complicated, non-local forms). Nevertheless, as we will show in the following sections, since the starting definitions are generally covariant, their actual values computed in the GLC gauge can be {\it used} in any other coordinate system (after expressing the GLC metric  in terms of the  coordinates and of the metric of the chosen system).

 We note that the expressions in Eqs. (\ref{areadist}) and (\ref{lumdist}) differ from the ones claimed in previous work \cite{BGMNV1, BGMNV2, BMNV, BGMNV3} by the presence of a factor $\det \left(u_{\tau}^{-1} \partial_\tau s_b^B \right)_{\la= \la_o}$ instead of a simple $\sin \tilde{\theta_o}$, the value it takes in the FLRW case. We shall now argue that this discrepancy can be eliminated by an appropriate choice of the angular coordinates along the observer's geodesic using the residual gauge freedom in Eq. (\ref{GLCres}). Indeed, we can always write:
\beq
\label{detvssin}
\det \left(u_{\tau}^{-1} \partial_\tau s_b^B \right)_{\la= \la_o}= \tilde{C}(\tau, \tilde{\theta}^a) \sin \tilde{\theta}_o \equiv C(w, \tilde{\theta}^a)  \sin \tilde{\theta}_o\, ,
\eeq
where  the correction, being computed on the observer's world-line, depends only on three variables: the angles and the proper time $\tau$ or, equivalently, the angles and the light-cone label $w$. Under the $\tau$-independent transformation  in Eq. (\ref{GLCres}):
\beq
\label{trSG}
\det \left(u_{\tau}^{-1} \partial_\tau s_b^B \right)_{\la= \la_o} \ra \det \left(\frac{\pa \bar{\theta}^a}{\pa \tilde{\theta}^b}\right)
 \bar{C}(w, \bar{\theta}^a)  \sin \bar{\theta}_o\, ,
\eeq
where we noticed that the operator $u_{\tau}^{-1} \partial_\tau$  transforms trivially  under  Eq. (\ref{GLCres}). The same result  follows from the invariance of $d_A^2$ under Eq. (\ref{GLCres}), implying that  both numerator and denominator in Eq. (\ref{areadist}) must change by the same  factor.  Given the way $\sqrt{\gamma(\la_s)}$ transforms, and that the angles are the same at the source and  the observer, the transformation in Eq. (\ref{trSG}) follows.
It is clear that, by an appropriate choice of the residual gauge transformation,  the determinant of the transformation  can cancel {\it at all $w$} the factor $\bar{C}$ and  recover the result $ \sin \bar{\theta}_o$ claimed in \cite{BGMNV1, BGMNV2, BMNV, BGMNV3}. 

In the following sections  we will express (in the simplest case of a static observer) the above result in terms of   perturbations in two more familiar gauges: the so-called synchronous and longitudinal (or Poisson) gauges, and compare the corresponding outcomes. In appendix B, we will also compute the luminosity distance $d_L$ as a function of the redshift, in the PG, up to second order in perturbation theory, and we will make connection with the existing literature.

%%%%%%%%%%%%%%%%%%%%%%%%%%%%%%%%%%%%%%%%%%%%%%%%%%%%%%%%%%%%%%%%%%%%%%%%%%%%%%%%%%%%%%%%
%%%%%%%%%%%%%%%%%%%%%%%%%%%%%%%%%%%%%%%%%%%%%%%%%%%%%%%%%%%%%%%%%%%%%%%%%%%%%%%%%%%%%%%%

%%%%%%%%%%%%%%%%%%%%%%%%%%%%%%%%%%%%%%%%%%%%%%%%%%%%%%%%%%%%%%%%%%%%%%%%%%%%%%%%%%%%
\section{First-order Jacobi map  in the synchronous gauge}
\label{Sec4}
\setcounter{equation}{0}
%%%%%%%%%%%%%%%%%%%%%%%%%%%%%%%%%%%%%%%%%%%%%%%%%%%%%%%%%%%%%%%%%%%%%%%%%%%%%%%%%%%%%%
As we have already stressed, the SG carries quite some affinity with the GLC gauge, in particular the two gauges share the same proper-time coordinate $\tau = t$ \cite{BGMNV1}. It is therefore natural to first check the expression of the JM in the SG.

At first order, the perturbed FLRW metric written in the SG, in the absence of vector and tensor perturbations, takes the well-known form (we use a bar to distinguish the SG scalar perturbations from those of the PG used in the next section):
\be
ds_{SG}^2 = - dt^2 + a^2(t) \left[ (1- 2\,\bar{\psi})\delta_{ij} + D_{ij} \bar{E}\right]  dx^i dx^j~~; ~~~~~~  D_{ij} = \partial_i \partial_j - \frac13 \delta_{ij} \Delta_3 \; ,
\label{SGmetricstandard}
\ee
where $\Delta_3 =  \nabla^2$ is the usual Laplacian operator in  3-dimensional Euclidean space.
In order to connect the SG perturbations to those of the GLC gauge we first transform the metric in Eq. (\ref{SGmetricstandard}) to standard spherical coordinates ($r, \theta, \phi$), writing it in the form:
\be
ds_{SG}^2 = - dt^2 + a^2(t) \left[ (1- 2\,Z) dr^2 - 2 S_a dr d\theta^a + h_{ab} d\theta^a d\theta^b \right] .
\label{SGspherical}
\ee
A direct calculation then gives
\bea
Z &=&   \bar{\psi} - \frac12 \left( \frac{\partial^2}{\partial r^2} - \frac13 \Delta_3 \right) \bar{E} ~~;~~~~~~~~  
S_a = - \left( \partial_r - \frac{1}{r} \right) \partial_a \bar{E} ,
 \nonumber \\ 
  h_{ab} &=&  \gamma^{0}_{ab} \left[1  - 2 \bar{\psi} - \left(  \frac13 \Delta_3 - \frac{1}{r} \partial_r\right) \bar{E} \right] + \nabla_a \partial_b \bar{E} ,
\eea
where  $\gamma^0_{ab}=r^{2}\,\text{diag}(1,\sin^{2}\theta) $ and $\nabla_a$ represents the covariant angular derivative.
In matrix form we then have
\be
g^{\mu\nu}_{SG} = 
\left(
\begin{array}{ccc}
-1& 0 & 0 \\ 0 & a^{-2} (1+2\,Z) & a^{-2} \gamma_0^{ab} S_b \\
0 & a^{-2}  \gamma_0^{ab} S_b & a^{-2}  h^{ab}
\end{array}
\right) \; ,
\ee
where  $\gamma_0^{ab}=r^{-2}\,\text{diag}(1,\sin^{-2}\theta) $ and
\be
h^{ab} = \gamma_0^{ab} \left[1  + 2 \bar{\psi}  +\left( \frac13 \Delta_3 -  \frac{1}{r} \partial_r\right) \bar{E} \right] 
- \gamma_0^{ac}  \gamma_0^{bd} \nabla_c \partial_d \bar{E} \, .
\ee

It is now quite straightforward to find the first-order coordinate transformation connecting the SG to the GLC gauge, and to derive the relation between  the corresponding first-order perturbations.
We introduce the useful (zeroth-order) light-cone variables $\eta_\pm= \eta \pm r$, 
with corresponding partial derivatives:
\beq
\pa_\eta = \pa_+ + \pa_- ~~~,~~~~~ \pa_r = \pa_+ - \pa_- ~~~,~~~~~\pa_\pm= {\pa \over \pa \eta_\pm}={1\over 2} \left( \pa_\eta \pm \pa_r \right) ~~,
\eeq
and impose the boundary conditions that $i)$ the transformation is non singular around $r=0$, and that $ii)$  the two-dimensional spatial section $r=$ const are locally parametrized at the observer position (for any $t$) by standard spherical coordinates. In this way
we find\footnote{These transformations were found in unpublished work done in collaboration with I. Ben Dayan and F. Nugier.}:
\bea
\label{GLCSG}
\tau &=& t, \\
w &=& r+\eta+\frac{1}{2}\int_{\eta_+}^{\eta_-}dx\, Z(\eta_+,x,\theta^a),  \\
\tilde{\theta}^a &=&  \theta^a + \frac12 \int_{\eta_+}^{\eta_-}dx \,\chi^a(\eta_+, x, \theta^a)\,,
\label{GLCSG}
\eea
where: 
\be
\label{xia}
\chi^a = S^a+\frac{1}{2}\gamma_0^{ac}\int_{\eta_+}^{\eta_-}dx\,\partial_c Z(\eta_+ ,x, \theta^a).
\ee
We then get
\bea
\Upsilon &=& a(\eta) \left[ 1-\frac{1}{2}(\partial_+ + \partial_-) \int_{\eta_+}^{\eta_-}dx Z(\eta_+,x,\theta^a) \right]\,,  \\
U^a &=&\frac{1}{2}(\partial_+ + \partial_-)  \int_{\eta_+}^{\eta_-}dx\, \chi^a(\eta_+,y,\theta^a)\,, \label{GLCtoSG} \\
a^2\gamma^{ab} &=& h^{ab}+\left[ \frac{1}{2}\gamma_0^{ac}\int_{\eta_+}^{\eta_-}dx\,\partial_c \chi^b(\eta_+ ,x, \theta^a)+a \leftrightarrow b 
\right]\,,
\eea
and we also have the following useful relation
\be 
\gamma^{-1} \equiv \det \gamma^{ab} = (a^2 r^2\sin\theta)^{-2} \left[ 4 \bar{\psi}-\frac{1}{3}\left(\Delta_3 -3 \partial_r^2\right)\bar{E}
+\partial_a \int_{\eta_+}^{\eta_-}dx \,\chi^a(\eta_+, x, \theta^a) \right].
\label{gammam1SG}
\ee

As already mentioned,  we assume our observer to be static in the SG, so that his/her worldline can be identified with $r =0$ by construction. 
Indeed, $u_{\mu} = - \partial_{\mu} \tau = - \partial_{\mu} t$ implies $ \dot{x}^{\mu}_{SG} \equiv u^{\nu} \partial_{\nu}  x^{\mu}_{SG} = \delta^{\mu}_0$. This velocity field also gives
\be
 \dot{x}^{\mu}_{GLC} =  (1, \Upsilon^{-1}, \Upsilon^{-1} U^a ) \; .
 \ee 
Furthermore,  let us show that, by having appropriately chosen the lower limits of integration in Eq. (\ref{GLCtoSG}), we were able to impose $\tilde{\theta}^a \rightarrow  \theta^a$ for $r\rightarrow 0$. 
 
To this purpose note first that the quantity $\chi^a$ of Eq. (\ref{xia}) remains finite as $r \rightarrow 0$. 
To see this, and for later use, let's expand $\bar{E}$ around the observer position. Choosing for simplicity  $\eta_o =0$, we obtain, up to second order,
\beq
\bar{E}(r,\eta, \theta)= E_0+E' \eta + E_i x^i+ {1\over 2} E^{\prime\prime} \eta^2 +{1\over 2} E_{ij} x^i x^j + E_i' x^i \eta + \dots
\label{13}
\eeq
where $x^i=(r \sin \theta \cos \phi, r \sin \theta \sin \phi, r \cos \theta)$ and all coefficients are constant.
Using Eq.(\ref{13}), and applying the useful relations $\Delta_2 x^i=-2 x^i$ and $\Delta_2 (x^i x^j)=2 r^2 \delta^{i j}-6 x^i x^j$, where $\Delta_2$ is the 2-dimensional angular Laplacian related to $\Delta_3$ by
\beq
\Delta_3 = \frac{1}{r^2} \Delta_2  + (\partial_r^2 + \frac2r\partial_r   ) \,,
\label{Delta2}
\eeq
we obtain,  for $r \rightarrow 0$:
\be
S^a = -\frac{1}{2}\frac{\gamma_0^{ab}}{r} \pa_b \left[E_{ij}x^i x^j\right] + C_1~~,~~~~~~~~~~\frac{1}{2}\gamma_0^{ab}\int_{\eta_+}^{\eta_-}dx\,\partial_b Z(\eta_+ ,x, \theta^a)  = \frac{1}{2}\frac{\gamma_0^{ab}}{r} \pa_b \left[E_{ij}x^i x^j\right] +C_2 \,.
\ee
Here $C_1$ and $C_2$ are $r$-independent constant terms, and  we have used that $\partial_a \bar{\psi} \sim {\cal O}(r)$ for $r \rightarrow 0$. 
Therefore the integrand  in Eq. (\ref{GLCSG}) is finite at $r=0$ and, since the integration range goes to zero, it follows  that $\tilde{\theta}^a \rightarrow  \theta^a$. This also implies, through Eq. (\ref{GLCtoSG}), that $U^a =0 = \dot{\tilde{\theta}}^a$ on the observer's worldline, consistently with the fact that the angles coincide in the two coordinate systems (and that those of the synchronous gauge are obviously constant).

We now compute $d_A$ by applying Eqs. (\ref{areadist}) and (\ref{ExpDetS2a}) for the free-falling static observer of the SG. 
Considering   that $\gamma^{3/2}$ goes like ${\cal O} \left( r^6\right)$ for $r \rightarrow 0$, the only non zero contribution to  Eq.(\ref{ExpDetS2a}) will be the one obtained by applying $\partial_\tau$  to the terms in $\gamma^{ab}$ that are  ${\cal O} \left(r^{-2}\right)$ for  $r \rightarrow 0$.  
Defining $\hat{\gamma}_{ab} = r^{-2} \gamma_{ab}$ we find:
\be
\label{den}
\det \left( \partial_\tau s_b^B(\lambda_o) \right)  = \left( \frac{\pa r}{\pa \tau}\right)^2_o \hat{\gamma}_o^{1/2} ~~,~~~~~~~~~~  \hat{\gamma}_o \equiv \det(\hat{\gamma}_{ab})_o \,\,.
\ee
We thus need  to evaluate $\hat{\gamma}_o$ and the $\tau$-derivative of $r$ at the origin. This can be done by inverting the $2 \times 2$ matrix of the derivatives of the GLC coordinates $\tau$ and $w$ with respect to $\eta$ and $r$
(being $\tilde{\theta}^a \equiv \theta^a$ at the origin, the further contributions coming from inverting the full $4 \times 4$ Jacobian matrix is negligible for our purpose). 
The result is
\be
\label{drdtau}
 \left( \frac{\pa r}{\pa \tau}\right)^2_o = a^{-2}_o ( 1 + 2 Z_o) = a^{-2}_o \left[ 1 + 2  \bar{\psi}_o  + \left(\frac{1}{3 r^2} \Delta_2 + \frac{2}{3r} \partial_r - \frac{2}{3} \partial_r^2\right) \bar{E}_o\right] \,, 
\ee 
 and we also have
\be
  \hat{\gamma}^{1/2}_o  = a_o^2 \sin \theta_o \left[ 1 - 2 \bar{\psi}_o +\frac13 \left(   \frac{ \Delta_2}{2r^2}    
 -   \partial_r^2 +  \frac{1}{r} \partial_r \right) \bar{E}_o  \right] \,,
\ee
where we have used Eqs.(\ref{gammam1SG}) and (\ref{Delta2}).
We then easily find:
\bea
\label{SGresult}
\frac{  \sin \theta_o}{\det \left( \partial_\tau s_b^B(\la_o) \right)  }&=&1- \frac12 \left( \Delta_3 - 3 \partial_r^2 \right) \bar{E}_o = 1-  \frac12 \left(\frac{ \Delta_2}{r^2} - 2  \partial_r^2 + 2 \frac{1}{r} \partial_r \right)  \bar{E}_o \,.
\eea

The above general expression appears to imply a correction to the simplest formula $d_A^2=\sqrt{\gamma_s}/\sin \theta_o$, used e.g. in \cite{BGMNV3}\footnote{However, such a possible correction trivially disappears if one computes (as in \cite{BGMNV3}) the angular average of the flux.}. 
In fact, by using the expansion in Eq. (\ref{13}), we obtain the following $r$-independent result\footnote{We stress that $x^i$-independent or linear  terms in the $x^i$  do not contribute to any physical quantity in the SG since the SG metric itself is insensitive to such terms. Furthermore, higher-order terms in the expansion of Eq. (\ref{13}) also give vanishing contributions at the observer.}
\begin{eqnarray}
\label{SGresultfinal}
\det \left( \partial_\tau s_b^B(\la_o) \right) &=&  \sin \theta_o \left[1 + \frac{1}{2r^2} \Delta_2 \left(\frac{1}{2} E_{ij} x^i x^j\right)\right]
\nonumber \\
&=& \sin \theta_o \left[1 + \frac{1}{2}\left(\delta^{ij} E_{ij}-\frac{3}{r^2} E_{ij} x^i x^j\right)\right]~,
\end{eqnarray}
instead of the simple result $ \sin \theta_o$. 

As in the case of the GLC gauge, we can try to use the residual gauge symmetry of the SG to remove the correction. Taking indeed the time-independent coordinate transformation:
\be
\label{SGCT}
x^i \ra \bar{x}^i = x^i  + \frac12 L^i_{j} x^j \, ,
\ee
we can choose the constant matrix $L^i_{j}$ as to make the new $D_{ij}\bar{E}$  vanish {\it at a given time}. This  can be chosen to be the present observer's time $\eta_o = 0$, which clearly corresponds to setting $E_{ij} =0$. Therefore, the gauge freedom in the SG is weaker than the one in the GLC, where we were able to remove the correction all along the observer's world-line.
Actually, this is what we should expect since the GLC transformation involves a change both in $\gamma_{ab}$ and in $U^a$. Hence, in general, this residual gauge fixing implies non-vanishing $U^a$ and thus prevents identifying the GLC angles with those in the SG all along the observer's geodesic. This is why in the SG we cannot remove the correction at all times. 

We finally combine together numerator and denominator of Eq. (\ref{areadist}), and  get for the area distance  the following SG result:
\bea
&&
\left(d_A^2 \right)_{SG}={ a_s^2 r_s^2 \sin \theta_s
\over \sin \theta_o} \Bigg\{1+\left({3\over 2} \pa_r^2 -{1\over 2} \Delta_3 \right) \bar{E}_o
-2 \bar{\psi}_s +\left({1\over 6} \Delta_3 -{1\over 2} \pa_r^2 \right) \bar{E}_s 
\nonumber \\
\Bigg.
&&
+{1\over 2}\!\int_{\eta_s^+}^{\eta_s^-}dx \ga_0^{ab}\pa_a\pa_b\!\left(\pa_r\!-\!{1\over r}\right) \bar{E}
\!-\!{1\over 4}\!\int_{\eta_s^+}^{\eta_s^-} dx \ga_0^{ab}\int_{\eta_s^+}^x dy\pa_a\pa_b\!\left[\bar{\psi} +\left({1\over 6} 
\Delta_3\!-\!{1\over 2} \pa_r^2\right) \bar{E} \right] \Bigg\}.
\label{d_ASG}
\eea
Actually,  previous  calculations of $d_A$ in the SG are available in the literature. 
Namely, in \cite{Barausse:2005nf} the area (and luminosity) distance  is given in the SG up to second order,  but only for 
a dust-dominated Universe. So far, we were not able to prove that our general first-order result agrees with theirs 
up to a residual gauge fixing.

%%%%%%%%%%%%%%%%%%%%%%%%%%%%%%%%%%%%%%%%%%%%%%%%%%%%%%%%%%%%%%%%%%%%%%%%%%%%%%%%%%%%%%
\section{First- (and second-) order Jacobi map in the Poisson gauge}
\label{Sec5}
\setcounter{equation}{0}
%%%%%%%%%%%%%%%%%%%%%%%%%%%%%%%%%%%%%%%%%%%%%%%%%%%%%%%%%%%%%%%%%%%%%%%%%%%%%%%%%%%%%%
Let us now consider another interesting gauge where our free-falling observer is no longer static:  the so-called Poisson gauge (PG) \cite{PG},  which represents a generalization of the Newtonian (or longitudinal) gauge beyond first order.  Neglecting  vector and tensor contributions (see \cite{BGMNV3}  for the motivations), the PG metric 
takes the following  form:
\bea
ds_{PG}^2 &=& a^2(\eta) \left[ -(1+ 2 \Phi) d\eta^2  + (1- 2 \Psi)\delta_{ij}  dx^i dx^j \right]  \nonumber \\
&=&  a^2(\eta) \left[ -(1+ 2 \Phi) d\eta^2  + (1- 2 \Psi)( dr^2 + r^2 d^2 \Omega) \right] 
\label{PGmetricstandard}
\eea
where $\Phi$ and $\Psi$ are scalar perturbations defined, up to second order, as:
\be
\Phi \equiv \phi + \frac{1}{2} \phi^{(2)} ~~,~~~~~~~ \Psi \equiv \psi + \frac{1}{2} \psi^{(2)} ~~,
\ee
and, in principle, we don't need to make any assumption about their possible dynamical sources. 

In order to compute the area distance in terms of standard PG perturbations we have to transform the GLC gauge quantities appearing in  Eq. (\ref{areadist}) to quantities of the PG. 
This has already been done in \ \cite{BGMNV1, BMNV} for the particular case of no-anisotropic stress (where $\phi = \psi$). Here we  extend the analysis to the more general case (but only for the first order) and take care of a small, inconsistency present in those analysis.
The point is that we have to impose suitable boundary conditions and, as in the previous section, we impose that $i)$ the transformation is non singular around $r=0$, and that $ii)$  the two-dimensional spatial section $r=$ const are locally parametrized at the observer position by standard spherical coordinates.
However,   unlike the case of the SG, for the PG these conditions  can only be imposed at the observer's space-time position (defined as $\eta=\eta_o$ and $r=0$) since, as a consequence of the dynamical motion of the PG free-falling observer, the observer is no longer at the origin $(r=0)$ of our coordinates system  for $\eta \ne \eta_o$.  

By considering such a physical property of the PG we obtain slightly different results with respect to those in \cite{BGMNV1, BMNV}~\footnote{In practice, all the integrals between $\eta_+$ and $\eta_-$, present in \cite{BGMNV1, BMNV}, become integrals between $\eta_o$ and $\eta_-$. The same applies to the vector and tensor perturbations considered in \cite{BMNV}.}, where such condition was imposed improperly at any $\eta$. To first order, in particular, we find (see Appendix B for the full second order transformation):
\bea
\tau &=& \tau^{(0)}+\tau^{(1)} =  \int_{\eta_{in}}^\eta d\eta' a(\eta')  + a(\eta) P(\eta, r, \theta^a) ~,
 \label{tauText} \\
w &=& w^{(0)}+w^{(1)} = \eta_+ + Q(\eta_+, \eta_-, \theta^a)  ~~,
  \label{wText} \\
\tilde{\theta}^a &=& \tilde{\theta}^{a (0)}+\tilde{\theta}^{a (1)} = \theta^a + \frac12 \int_{\eta_o}^{\eta_-} dx~ \left[ \gamma_0^{ab} \partial_b Q \right] (\eta_+,x,\theta^a)  ~~,
\label{thetatilde1orderShort}
\eea
 where we have defined:
\be
P(\eta, r, \theta^a) = \int_{\eta_{in}}^\eta d\eta' \frac{a(\eta')}{a(\eta)} \phi(\eta',r,\theta^a)
\,\,\,\,,\,\,\,\, ~~~ Q(\eta_+, \eta_-, \theta^a) = \int_{\eta_o}^{\eta_-} dx~ \frac{1}{2}\left(\psi + \phi \right)(\eta_+,x,\theta^a) ~,
\label{PQ}
\ee
and where the superscripts $(0), (1)$ denote, respectively, the background and first-order values of the given quantity. 
Finally, $\eta_{in}$ represents an early enough time when the perturbation (or better the integrand) was negligible: this means that the integrals over all relevant perturbation scales are insensitive to the actual value of $\eta_{in}$. To first order we can then use  
Eqs. (\ref{tauText}), (\ref{wText}) and (\ref{thetatilde1orderShort}) 
to compute the non-trivial entries of the GLC metric of Eq. (\ref{GLCmetric}), and obtain:
\bea
\Upsilon^{-1} &=& \frac{1}{a(\eta)} \left( 1 + \partial_+ Q - \partial_r P  \right) \,,
\label{Ups1}  
\\
U^a &=& \partial_{\eta}\tilde{\theta}^{a (1)}-\frac{1}{a}\gamma_0^{ab}\partial_b \tau^{(1)},
\label{Ua1}
\\ 
\gamma^{ab} &=& a^{-2} \left\{\gamma_0^{ab} \left(1 +  2 \psi\right) +\left[\gamma_0^{a c} \partial_c \tilde{\theta}^{b (1)}+ (a\leftrightarrow b) \right]\right\}.
\label{gammaabord1}
\eea

We now follow the same procedure as in the previous section. On one hand we need to evaluate the determinant of $\gamma$. From Eq. (\ref{gammaabord1}) we obtain:
\begin{equation}
\gamma^{-1} \equiv \det \gamma^{ab} =
(a^2r^2\sin\theta)^{-2} \left[1 + 4 \psi + 2 \partial_a \tilde{\theta}^{a (1)}\right].
\label{detord1}
\end{equation}
The other missing quantity is $\left(\partial r/\partial \tau\right)_{r=0}$, which can be evaluated by exactly the same method as the one followed in Sect. 4. At first order we  find:
\begin{equation}
\label{drdtaupg}
\left(\frac{\partial r}{\partial \tau} \right)_{r=0} = -\frac{1}{a(\eta_o)} \left(1+\psi_o+\partial_r P_o \right)\,.
\end{equation}
Combining Eqs. (\ref{detord1}) and (\ref{drdtaupg})  we arrive at the simple result:
\be
\label{sinpg1}
 \frac{  \sin \theta_o }{\det \left( \partial_\tau s_b^B(\la_o) \right)  }  =
1 - 2 \pa_r P_o ~.
\ee
We finally combine together numerator and denominator in Eq. (\ref{areadist}) to get:
\bea
\left(d_A^2 \right)_{PG} &=& {a_s^2 r_s^2 \sin \theta_s
\over \sin \theta_o} \left\{1- 2 \int_{\eta_{in}}^{\eta_o} d \eta' \frac{a(\eta')}{a(\eta_o)} \partial_r\phi\left(\eta', 0, \theta^a \right)-
 \right. \nonumber \\
 & & \left.
- 2 \psi_s -
{1\over 2} \int_{\eta_o}^{\eta_s^-} dx \gamma_0^{ab}\int_{\eta_o}^x dy \frac{1}{2} \pa_a\pa_b 
\left[\psi(\eta_+, y, \theta^a)+\phi(\eta_+, y, \theta^a)\right] \right\}\,.
\label{d_ALG}
\eea
This result will be compared to the corresponding one in the SG (see Eq. (\ref{d_ASG})) in Sect. 6.

Restricting now  ourselves to the case  of no anisotropic stress we set $\phi=\psi$. Furthermore, using the results given in Appendix B  where the coordinate transformation between the GLC gauge and the PG is given up to second order in perturbation theory for this case, we can generalize the previous results for $d_A$. In particular, Eq. (\ref{drdtaupg}) becomes: 
\begin{eqnarray}
\label{drdtaupg2}
\left(\frac{\partial r}{\partial \tau} \right)_{r=0} &=& -\frac{1}{a_o} \left\{1+\psi_o+\partial_r P_o+\frac{3}{2} \psi_o^2+2 \psi_o \partial_r P_o
+\frac{1}{2} \left(\partial_r P_o\right)^2+\frac{1}{2} \psi^{(2)}_o \right. \nonumber \\
& & \left.
+\frac{1}{2}  \int_{\eta_{in}}^{\eta_o} d \eta' \frac{a(\eta')}{a(\eta_o)} \partial_r \left[\phi^{(2)} - \psi^2 + \left(\partial_r P\right)^2 + \gamma_0^{ab} \partial_a P  \partial_b  P    \right]\left(\eta',0,\theta^a\right) \right\}.
\end{eqnarray}
Considering also the expression for $\ga$ generalized up to second order (see Eq.(\ref{gammaab}) in Appendix B), after a straightforward  calculation we obtain 
\bea
\label{sinpg2}
 \frac{  \sin \theta_o }{\det \left( \partial_\tau s_b^B(\la_o) \right)}  &=&
1+\Big[ - 2 \pa_r P + 2 (\pa_r P)^2 + \gamma_0^{ab} \partial_a P  \partial_b  P  - 2 \psi \pa_r P \Big]_o\nonumber   \\
&-& 
   \int_{\eta_{in}}^{\eta_o} d \eta' \frac{a(\eta')}{a(\eta_o)}\pa_r \left [\phi^{(2)} - \psi^2 +( \pa_r P)^2 + \gamma_0^{ab} \partial_a P  \partial_b  P    \right] ~.
\eea
This result can also be rewritten in the convenient form:
\beq
\label{final}
 \frac{  \sin \theta_o }{\det \left( \partial_\tau s_b^B(\la_o) \right)} = 
\Big[ (1 - \pa_r P)^2 + \nabla_i P  \nabla^i P  - 2 \psi_o \pa_r P\Big]_o - \int_{\eta_{in}}^{\eta_o} d \eta' \frac{a(\eta')}{a(\eta_o)}\pa_r \left(\phi^{(2)} - \psi^2 +  \nabla_i P  \nabla^i P   \right),
\eeq
where $\nabla_i$ is the gradient operator in polar coordinates. 

This contribution to $d_A^2$ can be matched, following \cite{SEF} and Sect. 3, to the kinematic correction $d \Omega /d \widetilde{\Omega}$, where $d \Omega$ is the infinitesimal solid angle measured at $o$ by our free-falling observer, while  $d \widetilde \Omega$ is the infinitesimal solid angle measured at $o$ by another observer which is static in the considered gauge.
Indeed, considering that our observer has a peculiar velocity $\vec{v}$, we find that the infinitesimal solid angle $d \Omega$ transforms under a Lorentz boost as follows: 
\beq
\label{solidangle}
d \Omega=\frac{1 - v^2}{(1- \vec{v}\cdot \vec {n})^2}\, d \widetilde{\Omega},
\eeq
where $\vec n$ is the unit vector along the direction connecting  the source to the observer. We then expect that the effect of the velocity on $d_A^2$ can be factorized (up to second order) as:
\beq
d_A^2 = \left[\frac{(1- \vec{v}\cdot \vec {n})^2}{1- v^2}\right]_o\left(\frac{\sqrt{\gamma_s}}{\sin \theta_o} \right) =  \Big[1+ v^2-2 \vec{v}\cdot \vec {n} +(\vec{v}\cdot \vec {n})^2\Big]_o\left(\frac{\sqrt{\gamma_s}}{\sin \theta_o} \right)\,.
\label{dArel}
\eeq
On the other hand, since $\tau$  plays the role of the effective gauge-invariant velocity potential, we can expand the spatial components of the perturbed velocity $v_\mu$ of the PG (geodesic) observer as:
\begin{equation}
v_i=-\partial_i \tau^{(1)} -\partial_i \tau^{(2)}~,~~~~~~~~~
v^i=-\frac{1}{a^2} \left[\partial^i \tau^{(1)} + \partial^i \tau^{(2)}+2 \psi \partial^i \tau^{(1)} \right]\,,
\label{GenVel}
\end{equation}
where $ \tau^{(1)}$ and $ \tau^{(2)}$ denote, respectively, the first- and second-order part of the coordinate transformation $\tau= \tau (\eta, r, \theta^a)$ between PG and GLC gauge (see  Eq.(\ref{tau2order}) in Appendix B for their explicit expressions). 
We can also expand  the unit vector $n_\mu$, in polar coordinates and to first order (which is enough), as:
\begin{equation}
n^\mu=\left(0, -\frac{1}{a}(1+\psi), 0, 0\right)\,\,\,\,\,\,\,\,,\,\,\,\,\,\,\,\,n_\mu=\left(0, -a(1-\psi), 0, 0\right).
\end{equation}
Inserting the above expressions for $\vec{v}$ and $\vec{n}$ into 
Eq. (\ref{dArel}) we get, up to second order:
\begin{eqnarray}
&&
\!\!\!\!\!
\Big[1+ v^2-2 \vec{v}\cdot \vec {n} +(\vec{v}\cdot \vec {n})^2\Big]_o =\nonumber \\
   && =  \Big[(1 - \pa_r P)^2 + \nabla_i P  \nabla^i P  - 2 \psi \pa_r P\Big]_o - \int_{\eta_{in}}^{\eta_o} d \eta' \frac{a(\eta')}{a(\eta_o)}\pa_r \left(\phi^{(2)} - \psi^2 +  \nabla_i P  \nabla^i P   \right)\,.
%   \nonumber \\ &&
\label{dArel2}
\end{eqnarray}
This shows that the perturbative corrections to the background relation $\det ( \partial_\tau s_b^B(\lambda_o))= \sin \theta_o$, appearing in Eq. (\ref{final}), can be exactly interpreted (to second order) as the effect of the peculiar velocity of our free-falling observer in the PG.  More generally, such velocity coincides with  the so-called gauge-invariant velocity perturbation \cite{Mukhanov}.

As in the case of the SG, we may ask whether this peculiar-velocity effect can be removed by some further gauge fixing. The answer in this case is negative. First of all, there is no residual gauge symmetry in the PG \cite{Mukhanov}. Also, if we fix the GLC gauge in order to recover the uncorrected result (as explained in Sec. 3), the GLC angles will not coincide with those in the PG even at $\eta = \eta_o$ and  the peculiar velocity correction should now be a consequence of the modified coordinate transformation connecting the two gauges.

An interesting property of the kinematic correction in Eq. (\ref{solidangle}) is that, upon integration over the whole solid angle, the result one obtains is always $4 \pi$ {\it quite independently of the peculiar velocity $v$}.
Therefore, if we are only interested in  the averaged energy flux (i.e. in $\langle d_L^{-2} \rangle$) on constant-redshift surfaces, all perturbative contributions to Eq. (\ref{final}) (that were missed in  \cite{BGMNV2,BMNV, BGMNV3}) simply drop out! 
This is no longer true if one computes correlations functions or dispersions around averaged values, or in case one wants to average quantities other than the flux. In those cases the contributions of the velocity corrections are nonvanishing, but they turn out to be numerically subleading with respect to the other contributions already considered in \cite{BGMNV2,BGMNV3}.

In order to obtain the complete second-order expression of $d_A$ we need  finally to combine numerator and denominator of Eq. (\ref{areadist}), expressed in terms of  PG quantities. The result of this long, but straightforward calculation is reported in Appendix B, where the final form of the luminosity distance as a function of the redshift is presented.

%%%%%%%%%%%%%%%%%%%%%%%%%%%%%%%%%%%%%%%%%%%%%%%%%%%%%%%%%%%%%%%%%%%%%%%%%%%%%%%%%%%%%%%%%%%%%%%%%%%%%%%%%%%
%%%%%%%%%%%%%%%%%%%%%%%%%%%%%%%%%%%%%%%%%%%%%%%%%%%%%%%%%%%%%%%%%%%%%%%%%%%%%%%%%%%%%%%%%%%%%%%%%%%%%%%%%%%

%%%%%%%%%%%%%%%%%%%%%%%%%%%%%%%%%%%%%%%%%%%%%%%%%%%%%%%%%%%%%%%%%%%%%%%%%%%%%%%%%
\section{Comparison of  the results in two gauges}
\label{Sec6}
\setcounter{equation}{0}
%%%%%%%%%%%%%%%%%%%%%%%%%%%%%%%%%%%%%%%%%%%%%%%%%%%%%%%%%%%%%%%%%%%%%%%%%%%%%%%%%%%%%

In order to compare the previous results for $d_A$ obtained in two different gauges, and check their physical equivalence (up to first order), let us consider the infinitesimal gauge transformation which connects the SG to the (first-order) PG. 
An ``infinitesimal'' coordinate transformation can be  parameterized, to  first-order, by the generator  $\ep_{(1)}^\mu$ as (see, for example,
\cite{Mukhanov}):
\beq
x^\mu \rightarrow \tilde{x}^\mu= x^\mu + \epsilon^\mu_{(1)}, 
\label{311}
\eeq
where
\beq
\ep_{(1)}^\mu= \left( \ep_{(1)}^0, \pa^i \ep_{(1)}+ \ep_{(1)}^i 
\right)\,.
\label{312}
\eeq
Under the associated gauge transformation (or local field reparametrization) -- where, by definition, old and new fields are evaluated at the same space-time position -- a tensor object changes, to first order, as
\be
T^{(1)} \rightarrow \tilde{T}^{(1)}=  T^{(1)}-L_{\ep_{(1)}} T^{(0)},
\label{GGT1}
\ee
where $L_{\ep_{(1)}}$ is the Lie derivative performed with respect to the vector $\ep^\mu_{(1)}$. 
Following \cite{Marozzi} 
 we then obtain the following relations between SG and PG quantities:  
\bea
\phi &=& -{a' \bar{E}'\over 2a} -{\bar{E}''\over 2}\,, 
\label{2} \\
\psi &=& \bar{\psi} +{a' \bar{E}'\over 2a} +{1\over 6} \Delta_3 \bar{E},
\label{2b} \\
\left(d_A^2 \right)_{PG} &=& \left(d_A^2 \right)_{SG} -\left[\ep^\mu \pa_\mu \left(d_A^2 \right)^{(0)}\right]_o
-\left[\ep^\mu \pa_\mu \left(d_A^2 \right)^{(0)}\right]_s,
\label{4}
\eea
where the components of $\ep_{(1)}^\mu$ are given by~\footnote{We consider only scalar fluctuations and consequently put $\ep^i_{(1)}=0$.}
\beq
\ep^0_{(1)}= {a \bar{E}'\over 2}\,, ~~~~~~~~~~~~~~ \ep_{(1)}= {1\over 2} \bar{E}\,,  ~~~~~~~~~~~~~~ \ep^i_{(1)}=0\,,
\label{5}
\eeq
and where we have taken into account the fact that  $d_A^2$ is a bi-scalar object. 

By applying the above relations we can express the PG result completely in terms of the SG variables as 
\bea
\left(d_A^2 \right)_{PG} &=& {a_s^2 r_s^2 \sin \theta_s
\over \sin \theta_o} \left[1+\pa_r \bar{E}_o'-2 \bar{\psi}_s  -\frac{a_s'}{a_s} \bar{E}_s' -{1\over 3} \Delta_3 \bar{E}_s\right.
 \nonumber \\
& & \left.
 -{1\over 4} \int_{\eta_o}^{\eta_s^-} dx\ga_0^{ab}\int_{\eta_o}^x dy \pa_a\pa_b \left(\bar{\psi} +{1\over 6} \Delta_3 \bar{E} -
 {\bar{E}''\over 2} \right)
\right]\,.
\label{6}
\eea
This has to be compared with the r.h.s of Eq.(\ref{4}), which is given by
\bea
&& \!\!\!\!\!\!\!\!\!\!\!\!\!\!\!\!\left(d_A^2 \right)_{SG} -\left[\ep^\mu \pa_\mu \left(d_A^2 \right)^{(0)}\right]_o
-\left[\ep^\mu \pa_\mu \left(d_A^2 \right)^{(0)}\right]_s =
{a_s^2 r_s^2 \sin \theta_s
\over \sin \theta_o} \Bigg[1+\left({3\over 2} \pa_r^2 \bar{E} -{1\over 2} \Delta_3 \bar{E} \right)_o
 \nonumber \\
\Bigg.
&&
-2 \bar{\psi}_s +\left({1\over 6} \Delta_3-{1\over 2} \pa_r^2\right)\bar{E}_s+{1\over 2} \int_{\eta_s^+}^{\eta_s^-}dx \ga_0^{ab} \pa_a \pa_b \left(\pa_r -{1\over r}\right)\bar{E} \nonumber
\\ \nonumber 
\Bigg.
&&
-{1\over 4} \int_{\eta_s^+}^{\eta_s^-} dx \ga_0^{ab}\int_{\eta_s^+}^x dy \pa_a\pa_b \left(\bar{\psi} +{1\over 6} \Delta_3 \bar{E} 
- {1\over 2} \pa_r^2 \bar{E} \right)+\left(\frac{1}{2 r^2}\cot \theta \partial_\theta \bar{E}\right)_o
\\  
\Bigg.
&& -\frac{a_s'}{a_s}\bar{E}_s'-\frac{1}{r_s} \partial_r \bar{E}_s -\left(\frac{1}{2 r^2}\cot \theta \partial_\theta \bar{E}\right)_s
\Bigg].
\label{7}
\eea
At  first sight  Eq.(\ref{7}) has no contributions to $d_A^2$ from terms in $\bar{E}$ that are linear in $x^i$ when expanded around the observer's position. This is in clear contrast with the  PG expression of Eq.(\ref{6}) which, on the contrary, includes such terms (see, in particular, the contribution $(\pa_r \bar{E}')_o$ describing the peculiar velocity of the observer). Besides this, the two expressions appears to differ in many other respects. However, as we shall now see, the two expressions are essentially the same, apart from a small difference
that can be made to disappear (at a given value of the observer's time) by exploiting the residual gauge freedom of the SG. 

Indeed, by using in the integral of the third line of Eq.(\ref{7}) the identity $\pa_r^2 = \pa_\eta^2 - 4 \pa_- (\pa_\eta - \pa_-)$, we find that the $\pa_\eta^2$ contribution exactly reproduces the corresponding terms of the double integral of Eq. (\ref{6}). The remaining contributions from $\pa_r^2$ can be integrated once and added to the integral appearing in the second line of Eq. (\ref{7}) to give:
\bea
+{1\over 2} \int_{\eta_s^+}^{\eta_s^-} dx \ga_0^{ab} \pa_a \pa_b \left\{ \left(\pa_r -{1\over r} \right)\bar{E}(\eta_+, x, \theta^a) 
+\left(\pa_x -\pa_\eta \right)\left[ \bar{E}(\eta_+, x, \theta^a)- \bar{E}(\eta_+,\eta_+, \theta^a)\right]\right\}.
\label{8}
\eea
These integrals can be done explicitly, giving contributions both at the source and at the observer. 
In particular, we can split the total contribution in two parts: the first one,  given by
\beq
-{1\over 2}\left( \ga_0^{ab} \pa_a \pa_b \bar{E} \right)_s,
\label{9}
\eeq
 nicely combines with all remaining source terms of Eq. (\ref{7}) to reproduce all remaining source terms of Eq. (\ref{6}).
The second contribution, obtained when at least one lower boundary of the various integrals is considered, is  given by:
\beq
{1\over 2}\ga_0^{ab} \pa_a \pa_b \bar{E}(\eta_+, \eta_+, \theta^a)  + \left[ \left( r \ga_0^{ab}\right)_s - \left( r \ga_0^{ab}\right)_o \right]  \pa_a \pa_b \pa_+ \bar{E}(\eta_+, \eta_+, \theta^a) ~.
\label{11}
\eeq
where $\bar{E}(\eta_+, \eta_+, \theta^a)$ is the limiting value of $\bar{E}$ approaching the observer position along the 
light-cone. 
All these contributions have to be evaluated at the observer, except for the geometric prefactor $\left(r \ga_0^{ab}\right)_s$ of the second term, which is 
referred to the source position. 
In order to obtain a full agreement between the results for $d_A$ in the two gauges (see Eq.(\ref{4})) 
we should thus verify the  following equality:
\be
\left(\pa_r \bar{E}'\right)_o=
\left({3\over 2} \pa_r^2-{1\over 2} \Delta_3\right) \bar{E}_o +{1\over 2} \left(\frac{\Delta_2}{r^2} \bar{E} \right)_o
+ \left[ \left( r \ga_0^{ab}\right)_s - \left( r \ga_0^{ab}\right)_o \right] \left(  \pa_a \pa_b \pa_+ \bar{E}\right)_o \,,
\label{12}
\ee
where we recall that the second and third terms on the r.h.s. have to be evaluated along the light-cone, while the first term does not depend on the limiting process followed to arrive at the observer's position, see Eqs. (\ref{SGresult}) and (\ref{SGresultfinal}), and has a different physical origin.

It is convenient, at this point, to use the expansion of the SG variable $\bar{E}$ around the observer position. Note, however, that we have to set $E_i =0$ since otherwise the gauge transformation in Eq. (\ref{5}) becomes singular at $r=0$ ($\delta\theta \sim r^{-2} E_i x^j \ra \infty$). 
With this proviso, using Eq.(\ref{13}), the expansion of the term on the l.h.s. of Eq. (\ref{12}) gives:
\beq
\left(\pa_r \bar{E}' \right)_o =\left({E'_i x^i\over r}\right)_o\, ,
\label{14}
\eeq
while the expansion of all terms on the r.h.s. gives:
\bea
&&
{1\over 2}E_{ij} \left({3 x^ix^j\over r^2} -\da^{ij} \right)_o -
\left[{1\over r^2} \left( E_i' x^i \eta -{1\over 4} 
E_{ij} \Delta_2 x^ix^j\right) \right]_o
\nonumber \\ &&
+{1\over 2}\left[ \left({1\over r_s} -{1\over r} \right) \left(\pa_\theta^2 +{1\over \sin^2 \theta} \pa_\phi^2 \right)
 \left({E_{ij}\over r} x^ix^j+{E_i' x^i\over r} \eta + E_i' x^i \right) \right]_o \nonumber \\ 
&&
\!\!\!\!
= \left({E'_i x^i\over r}\right)_o   - {1\over 2}\left[  \left(\pa_\theta^2 +{1\over \sin^2 \theta} \pa_\phi^2 \right) \left(
E_{ij}  \frac{x^ix^j}{r^2} \right) \right]_o.
\label{15}
\eea
To obtain the last line we have used the explicit definition of $\Delta_2$ and the fact that the explicit time-dependent contributions to the above equation, proportional to $E_i' x^i \eta$, are generated by the expansion of terms approaching the observer along light-cone trajectories. For such terms we can safely replace $\eta$ with $-r$ (to first order), so that the last two contributions of the second line exactly cancel between themselves, while the time-dependent contribution of the first line exactly matches the result of Eq. (\ref{14}). The only mismatch consists of the last term appearing in Eq. (\ref{15}).

Let us finally exploit the already mentioned residual gauge freedom of the SG, considering the (time independent!) gauge transformation (valid close to the observer's position)
with generator 
\be
\epsilon_{(1)}^\mu=\left(0, \partial^k  (\frac{x^i x^j}{4} E_{ij}) \right), 
\label{generatorGF}
\ee
leading to
\bea
\bar{E}(r,\eta, \theta) &\ra& \widetilde E(r,\eta, \theta) =\bar{E}(r,\eta, \theta) -{1\over 2} x^i x^j E_{ij}\,,
\label{16} \\
\bar{\psi}(r,\eta, \theta) &\ra& \widetilde \psi(r,\eta, \theta) = \bar{\psi}(r,\eta, \theta)+\frac{1}{6} \delta^{ij} E_{ij}\,.
\eea
By applying such a (partial) gauge fixing we can set $E_{ij}$  to zero in the small-$x$,  small-$\eta$ expansion of $ \widetilde E(r,\eta, \theta)$ and thus eliminate all terms of Eq. (\ref{15}) except for the $E_i'$ contribution. Within this gauge choice we thus find  full agreement  for the expression of $d_A$  in the two gauges at one observer's time.

We recall that precisely the same residual gauge fixing was used (see end of Sect. 4)   to remove the anisotropy correction  in the SG at one specific  time, i.e. to 
 remove anisotropy  around the observer ($E_{ij} =0$  at $\eta = \eta_o$). Since the PG corresponds to choosing shear-free equal-time hypersurfaces, it is hardly surprising that such a residual gauge transformation is necessary in order to recover agreement between the two gauges.

%%%%%%%%%%%%%%%%%%%%%%%%%%%%%%%%%%%%%%%%%%%%%%%%%%%%%%%%%%%%%%%%%%%%%%%%%%%%%%%%%%%%%%%%%%%%%%%%%%%%%%%%%%%
%%%%%%%%%%%%%%%%%%%%%%%%%%%%%%%%%%%%%%%%%%%%%%%%%%%%%%%%%%%%%%%%%%%%%%%%%%%%%%%%%%%%%%%%%%%%%%%%%%%%%%%%%%%

%%%%%%%%%%%%%%%%%%%%%%%%%%%%%%%%%%%%%%%%%%%%%%%%%%%%%%%%%%%%%%%%%%%%%%%%%%%%%%
\section{Summary and conclusions}
\label{Sec7}
\setcounter{equation}{0}
%%%%%%%%%%%%%%%%%%%%%%%%%%%%%%%%%%%%%%%%%%%%%%%%%%%%%%%%%%%%%%%%%%%%%%%%%%%%%

Let us briefly summarize the main results of this work.

We have shown that the equation of geodesic deviation can be solved {\it exactly} in the GLC gauge provided the origin of the null geodesic bundle is identified with the origin of the GLC spherical coordinates. In principle, this can be done all along the observer's world-line. The main results of this paper are  given in Eqs. (\ref{finalJM}), (\ref{areadist}) and (\ref{lumdist}) and show how, in this gauge, Jacobi map, area  and luminosity distance all factorize as products of a quantity evaluated at the observer times a quantity evaluated at the source (such a factorization was already observed in \cite{GMNV} for what concerns the redshift).

The importance of having found an exact, non-perturbative expression for these quantities can be hardly overestimated. It can allow, for instance, for a non-perturbative approach to the backreaction problem (i.e. understanding how inhomogeneities may affect  cosmological observables), to the so-called redshift drift, or to the study of large scale structure via micro-lensing, one of the main aims of the Euclid mission \cite{Euclid}.  It can also help with precision determinations of the different power spectra describing CMB anisotropies and large-scale structure.

The main computational obstacle for carrying out this program appears to be the present lack of a convenient formulation of Einstein's (constraint and evolution) equations  in this gauge. Another, more technical obstacle is that we have implicitly  limited ourselves to the case of no caustics (see e.g. \cite{Maartens3}), although a generalization of the solution in the presence of caustics should be possible.  We plan to study both issues in the near future.

In the second part of the paper (Sects 4, 5, 6 and Appendix B) we have compared our exact result with the perturbative ones already known in other gauges. In particular, we could find the expression for the area (luminosity) distance to first order in cosmological perturbation theory in the synchronous gauge  and up to second order in the Poisson gauge.
Comparing the results in those two gauges reveals some subtleties about how so-called aberration effects are encoded in our general result. Furthermore, we found that agreement between results in these two gauges demands  (partial)  fixing of the residual gauge symmetry characteristic of the SG. Similarly, we recover the already proposed  GLC formula for the area/luminosity distance at the price of a further fixing of the GLC coordinates.

The last result of this work, also presented in Appendix B, consists in a re-derivation of  the luminosity-redshift relation to second order in the PG starting directly from the JM. The result thus obtained differs slightly for the one proposed in \cite{BMNV} (and used in \cite{BGMNV2}, \cite{BGMNV3}), basically because of a refined approximation on some ``integration constants" (or ``integration boundaries") and of aberration terms which are generated in the PG. The modifications do not affect in any significant way the physical results presented in those papers.

%%%%%%%%%%%%%%%%%%%%%%%%%%%%%%%%%%%%%%%%%%%%%%%%%%%%%%%%%%%%%%%
\section*{Acknowledgements}
%%%%%%%%%%%%%%%%%%%%%%%%%%%%%%%%%%%%%%%%%%%%%%%%%%%%%%%%%%%%%%%%%

We wish to thank Guillermo Ballesteros, Enea Di Dio, Ruth Durrer, George F. R. Ellis, Roy Maartens and Jean-Philippe Uzan for useful discussions. 
MG and GV would like to thank the hospitality and support of the  
D\'epartement de Physique Th\'eorique of the University of Geneva while most of this work has been carried out. 
GF would like to thank the University of Geneva and the Coll\`ege de France for  hospitality during the research that led to this paper.

GM is supported by the Marie Curie IEF, Project NeBRiC - ``Non-linear effects and backreaction in classical and quantum cosmology".

%%%%%%%%%%%%%%%%%%%%%%%%%%%%%%%%%%%%%%%%%%%%%%%%%%%%%%%%%%%%%%%%%%%%%%%%%%%%%%%%%%%%%%%%%%%%%%%%%%%%%%%%%%%
%%%%%%%%%%%%%%%%%%%%%%%%%%%%%%%%%%%%%%%%%%%%%%%%%%%%%%%%%%%%%%%%%%%%%%%%%%%%%%%%%%%%%%%%%%%%%%%%%%%%%%%%%%%

%%%%%%%%%%%%%%%%%%%%%%%%%%%%%%%%%%%%%%%%%%%%%%%%%%%%%%%%%%%
\begin{appendix}
\renewcommand{\theequation}{A.\arabic{equation}}
\setcounter{equation}{0}
\section*{Appendix A. Some useful relations}
%%%%%%%%%%%%%%%%%%%%%%%%%%%%%%%%%%%%%%%%%%%%%%%%%%%%%

 In this appendix we provide some useful technical details.

First of all we list all  the Christoffel symbols that follow from the metric  \eqref{LCmetric}:
\bea
&\Gamma_{\tau\tau}^\rho&=\frac{\partial_\tau \Upsilon}{\Upsilon} \delta_\tau^\rho \quad \quad \quad \quad,\quad \quad \quad \quad  \Gamma_{ab}^w =\frac{\partial_\tau \gamma_{ab}}{2\Upsilon}\quad \quad,\quad \quad
\Gamma_{\tau w}^w=\Gamma_{\tau a}^w=0\,,\nonumber\\
&\Gamma_{\tau w}^\tau&=\frac{1}{2}\left[ \frac{\partial_\tau \left( \Upsilon^2+U^2 \right)}{\Upsilon}-\partial_w\left( \Upsilon^2+U^2 \right)+\frac{U^a}{\Upsilon}\left( \partial_\tau U_a-\partial_a\Upsilon \right) \right]\,,\nonumber\\
&\Gamma_{\tau a}^\tau&=\frac{1}{2}\left[ \frac{\partial_\tau U_a}{\Upsilon}+\frac{\partial_a \Upsilon}{\Upsilon}-\frac{U^b}{\Upsilon}\,\partial_\tau \gamma_{ab} \right]\,,\nonumber\\
&\Gamma_{\tau w}^a&=\frac{1}{2}\,\gamma^{ab}\left[ \partial_b\Upsilon-\partial_\tau U_b \right]\quad \quad,\quad \quad
\Gamma_{\tau a}^b=\frac{1}{2} \,\gamma^{bc}\,\partial_\tau \gamma_{ac}\,,\nonumber\\
&\Gamma_{ww}^\tau&=\frac{1}{2}\left[ 2\,\partial_w \Upsilon +\partial_\tau \left( \Upsilon^2+U^2 \right)-\frac{\partial_w(\Upsilon^2+U^2)}{\Upsilon}+\frac{U^a}{\Upsilon}\left(2\,\partial_wU_a+\partial_a(\Upsilon^2+U^2)  \right)\right]\,,\nonumber\\
&\Gamma_{ww}^w&=\frac{1}{2}\left[2\,\frac{\partial_w\Upsilon}{\Upsilon}+\frac{\partial_\tau(\Upsilon^2+U^2)}{\Upsilon}\right]\,,\nonumber\\
&\Gamma_{ww}^a&=\frac{1}{2}\left[\frac{U^a}{\Upsilon}\left[
2\,\partial_w\Upsilon+\partial_\tau(\Upsilon^2+U^2)\right]-\gamma^{ab}\partial_b(\Upsilon^2+U^2)\right]\,,\nonumber\\
&\Gamma_{wa}^\tau&=\frac{1}{2}\left[ \partial_a\Upsilon-\partial_\tau U_a-\frac{\partial_a(\Upsilon^2+U^2)}{\Upsilon}-\frac{U^b}{\Upsilon}\left( \partial_w\gamma_{ab}-\partial_a U_b+\partial_b U_a \right) \right]\,,\nonumber\\
&\Gamma_{wa}^w&=\frac{1}{2\Upsilon}\left[ \partial_a\Upsilon-\partial_\tau U_a \right]\,,\nonumber\\
&\Gamma_{wa}^b&=\frac{1}{2}\left[ \frac{U^b}{\Upsilon}\left( \partial_a\Upsilon-\partial_\tau U_a \right)+\gamma^{bc}\left( \partial_w \gamma_{ac}-\partial_aU_c+\partial_cU_a \right) \right]\,, \nonumber 
\\
&\Gamma_{ab}^\tau&=\frac{1}{2}\left[ \partial_\tau \gamma_{ab}+\frac{1}{\Upsilon} \left( \partial_b U_a+\partial_w \gamma_{ab} \right)-\frac{U^c}{\Upsilon}\left( \partial_a\gamma_{bc}+\partial_b\gamma_{ca}-\partial_c\gamma_{ab} \right) \right]\,,\nonumber
\eea
\bea
\!\!\!\!\!\!\!\!\!\!\!\!\!\!\!\!\!\!\!\!\!\!\!\!\!\!\!\!\!\!\!\!\!\!\!\!\!\!\!\!\!\!\!\!\!\!\!\!\!\!\!\!\!\!\!\!\!\!\!\!\!\!\!\!\!\!\!\!\!\!\!\!\!\!\!\!\!\!\!\!&\Gamma_{ab}^c&=\frac{1}{2}\left[ \frac{U^c}{\Upsilon}\,\partial_c\gamma_{ab}+\gamma^{cd}\left( \partial_a\gamma_{bd}+\partial_b\gamma_{da}-\partial_d\gamma_{ab} \right) \right]\,.
\eea

Second, let us show how one can  always  impose Eq. \eqref{eq:SachsBasis4} on the zweibein field $s_A^a$. We first note that Eq.  \eqref{eq:SachsBasis4} can be written explicitly as:
\beq
\label{parexp}
  k^\nu\nabla_\nu s_A^a = 0 =  \dot{s}_A^a + \Gamma_{\tau b}^{a} s_A^{b} = \dot{s}_A^a + \frac12  \gamma^{a c} \dot{\gamma}_{ cb} s_A^{b} .
\eeq
Expressing  $\gamma_{ab}$ and $\gamma^{ab}$ in terms of  the $s_A^a$ we obtain, after some trivial algebra:
\beq
\label{parcond}
  \dot{s}_A^a s_{aB} -   \dot{s}_B^a s_{aA} = 0 = \epsilon^{AB}  \dot{s}_A^a s_{aB} \, .
\eeq
Hence we get a single condition, rather than the two conditions naively expected.

Suppose now we are starting with some arbitrary zweibeins which, instead of satisfying Eq. (\ref{parcond}) are characterized by:
\beq
\epsilon^{AB}  \dot{s}_A^a s_{aB} = X \ne 0\, .
\eeq
Recalling that the zweibeins are defined up to a local 2-dimensional rotation $\Lambda^A_B$ of the flat indices, we can always define 
the rotated zweibeins
$\tilde{s}_A^a = \Lambda_A^B s_B^a$ such that:
\beq
\epsilon^{AB}  \dot{\tilde{s}}_A^a\tilde{s}_{aB} =  X + \epsilon^{AB}  \dot{\Lambda}_A^C\Lambda_B^D \delta_{CD} = X - 2 \dot{\alpha} \,,
\eeq
where $\alpha$ is the local rotation angle.
Clearly, by solving the simple differential equation $2 \dot{\alpha} = X$, we can always get rid of $X$ and thus find a suitable set of parallel-transported zweibeins.

\end{appendix}

%%%%%%%%%%%%%%%%%%%%%%%%%%%%%%%%%%%%%%%%%%%%%%%%%%%%%%%%%%%%%%%%%%%%%%%%%%%%%%%%%%%%%%%%%%%%%%%%%%
%%%%%%%%%%%%%%%%%%%%%%%%%%%%%%%%%%%%%%%%%%%%%%%%%%%%%%%%%%%%%%%%%%%%%%%%%%%%%%%%%%%%%%%%%%%%%%%%%%%%

%%%%%%%%%%%%%%%%%%%%%%%%%%%%%%%%%%%%%%%%%%%%%%%%%%%%%%%%%%
\begin{appendix}
\renewcommand{\theequation}{B.\arabic{equation}}
\setcounter{equation}{0}
\section*{Appendix B.  The luminosity distance $d_L$ up to second order}
%%%%%%%%%%%%%%%%%%%%%%%%%%%%%%%%%%%%%%%%%%%%%%%%%%%%%%%%%%%%%

In this appendix we first  extend to second order in perturbation theory the coordinate transformation between the GLC and Poisson gauges (in the absence of sources with anisotropic stresses, i.e. with $\phi = \psi$). We then derive the luminosity distance/redshift relation in the latter gauge considering a geodesic observer and a geodesic source.

Starting from  the boundary condition introduced in Sec. 5, the coordinate transformation can be generalized to second order, with self-explanatory notations, 
as follows:
\bea
\!\tau \!&=&\! \tau^{(0)}+\tau^{(1)}+\tau^{(2)} \nonumber \\
\!&=&\! \left( \int_{\eta_{in}}^\eta d\eta' a(\eta') \right) + a(\eta) P(\eta, r, \theta^a) \nonumber \\
& & + \int_{\eta_{in}}^\eta d\eta' \frac{a(\eta')}{2} \left[ \phi^{(2)} - \psi^2 + ( \partial_r P )^2 + \gamma_0^{ab} ~ \partial_a P ~ \partial_b P \right] (\eta', r, \theta^a)~,
\label{tau2order} \\
\!w \!&=&\! w^{(0)}+w^{(1)}+w^{(2)} \nonumber \\
\!&=&\! \eta_+ + Q(\eta_+, \eta_-, \theta^a) + { \frac{1}{4} \int_{\eta_o}^{\eta_-} dx~ \left[ {\psi}^{(2)} + {\phi}^{(2)} + 4 {\psi} \partial_+ Q + {\gamma}_0^{ab} ~ \partial_a Q ~ \partial_b Q \right] (\eta_+, x, \theta^a)} ~~,
\label{w2order} \\
\!\tilde{\theta}^a \!&=&\! \tilde{\theta}^{a (0)}+\tilde{\theta}^{a (1)}+\tilde{\theta}^{a (2)} \nonumber \\
\!&=&\! \theta^a + { \frac12 \int_{\eta_o}^{\eta_-} dx~ \left[ {\gamma}_0^{ab} \partial_b Q \right] (\eta_+,x,\theta^a)} 
+ { \int_{\eta_o}^{\eta_-} dx~ \left[ {\gamma}_0^{ac} \zeta_c + {\psi} ~ \xi^a + \lambda^a \right] (\eta_+,x,\theta^a)} ~~,
\label{thetatilde2orderShort}
\eea
where we have used the following shorthand notations:
\bea
\zeta_c(\eta_+,x,\theta^a) &=& \frac12 \partial_c w^{(2)} (\eta_+,x,\theta^a)  \nonumber \\
&=& { \frac18 \int_{\eta_o}^x du~ \partial_c \left[ {\psi}^{(2)} + {\phi}^{(2)} + 4 {\psi} ~ \partial_+ Q + {\gamma}_0^{ef} ~ \partial_e Q ~ \partial_f Q \right] (\eta_+,u,\theta^a)} ~~,
\label{ExprChi}
\\
\xi^a(\eta_+,x,\theta^a) &=& \partial_+ \tilde{\theta}^{a(1)}(\eta_+,x,\theta^a) +2\partial_x\tilde{\theta}^{a(1)}(\eta_+,x,\theta^a) ~~ \nonumber \\
&=& { \partial_+ \left( \frac12 \int_{\eta_o}^x du~ [{\gamma}_0^{ac} \partial_c Q] (\eta_+,u,\theta^a) \right)} + [\gamma_0^{ac} \partial_c Q] (\eta_+,x,\theta^a) ~~,
\label{ExprXi}
\\
\lambda^a(\eta_+,x,\theta^a) &=& \partial_x \tilde{\theta}^{d(1)} (\eta_+,x,\theta^a)\left(\partial_d \tilde{\theta}^{a(1)} (\eta_+,x,\theta^a)-\delta_\chi^a  \partial_+ Q (\eta_+,x,\theta^a)\right) ~~ \nonumber \\
&=& \frac14 [ {\gamma}_0^{dc} ~ \partial_c Q ] (\eta_+,x,\theta^a) \left( { \int_{\eta_o}^x du ~ \partial_d \left[ {\gamma}_0^{ae} \partial_e Q \right] (\eta_+,u,\theta^a)} \right) \nonumber \\
& & 
- \frac12 \left[ \partial_+ Q ~ {\gamma}_0^{ab} \partial_b Q \right] (\eta_+,x,\theta^a) ~~. 
\label{ExprLamb}
\eea
We can then compute the non-trivial entries of the GLC metric of Eq. (\ref{GLCmetric}), and obtain:
\bea
\Upsilon^{-1} &=& \frac{1}{a(\eta)} \left[ 1 + \partial_+ Q - \partial_r P  
+\partial_{\eta}w^{(2)} + \frac{1}{a}(\partial_\eta - \partial_r) \tau^{(2)} - \psi \partial_{\eta}Q - \phi^{(2)} + 2\psi^2 \right. \nonumber \\
& & \left.
- \partial_r P \partial_r Q - 2 \psi \partial_r P - \gamma^{ab}_0 \partial_a P \partial_b Q\right] \,,
\label{Ups1}  
\\
U^a &=& \partial_{\eta}\tilde{\theta}^{a (1)}-\frac{1}{a}\gamma_0^{ab}\partial_b \tau^{(1)}+\partial_{\eta}\tilde{\theta}^{a (2)}-
\frac{1}{a}\gamma^{ab}_0\partial_b \tau^{(2)} - \frac{1}{a} \partial_r \tau^{(1)} \partial_r \tilde{\theta}^{a(1)}  \cr
& &  -\psi \left(\partial_\eta \tilde{\theta}^{a (1)}+\frac{2}{a}\gamma^{ab}_0 \partial_b \tau^{(1)} \right)-\frac{1}{a}\gamma_0^{cd}\partial_c \tau^{(1)} \partial_d \tilde{\theta}^{a (1)} \nonumber \\
& &
+\left(\partial_+ Q - \partial_r P\right) \left(-\partial_{\eta} \tilde{\theta}^{a (1)}+\frac{1}{a}\gamma^{ab}_0 \partial_b \tau^{(1)}\right)\,,
\label{Ua1}
\\ 
\gamma^{ab} &=& a^{-2}\left\{ \gamma_0^{ab} \left(1 +  2 \psi\right) +\left[\gamma_0^{a c} \partial_c \tilde{\theta}^{b (1)}+ (a\leftrightarrow b) \right]+ \gamma_0^{ab}
\left(\psi^{(2)} + 4 \psi^2 \right)-\partial_\eta \tilde{\theta}^{a (1)}\partial_\eta \tilde{\theta}^{b (1)}
\right. \nonumber \\
& & \left.
+\partial_r \tilde{\theta}^{a (1)}\partial_r \tilde{\theta}^{b (1)} +2 \psi \left[\gamma_0^{a c} \partial_c \tilde{\theta}^{b (1)}+ (a\leftrightarrow b) \right]+\gamma_0^{c d} \partial_c \tilde{\theta}^{a (1)}
 \partial_d \tilde{\theta}^{b (1)} \right.
 \nonumber \\
& &  \left.
 +\left[\gamma_0^{a c} \partial_c \tilde{\theta}^{b (2)}+ (a\leftrightarrow b) \right] \right\}.
\label{gammaab}
\eea
We also easily obtain the useful relation:
\bea
\gamma^{-1} \equiv \det \gamma^{ab} &=& 
(a^2 r^2\sin\theta)^{-2} \left\{1 + 4 \psi + 2 \partial_a \tilde{\theta}^{a (1)} +2 \psi^{(2)}
+12 \psi^2+ 2 \partial_a \tilde{\theta}^{a (2)}
\right. \nonumber \\
& & \left.
\!\!\!\!\!\!\!\!-4 \gamma_{0 a b} \partial_+ \tilde{\theta}^{a (1)}\partial_- \tilde{\theta}^{b (1)}
+ ~ 8 \psi \partial_a \tilde{\theta}^{a (1)}+2 \partial_a \tilde{\theta}^{a (1)}
\partial_b \tilde{\theta}^{b (1)}
-\partial_a \tilde{\theta}^{b (1)}\partial_b \tilde{\theta}^{a (1)}
\right\}\!.
\label{det}
\eea

We start from  Eq. (\ref{lumdist})  and use Eqs. (\ref{final}) and (\ref{det}). Then,  following \cite{BMNV}, we obtain
the final expression of the luminosity distance in terms of  perturbations in the PG, of the observed redshift, and of the observer's angular coordinates $\tilde{\theta}^a$.

We underline that the result to be obtained is valid in general, i.e. without the need of considering a particular (e.g. CDM or  $\Lambda$CDM)
cosmology.
It is also an improvement with respect to the result given in \cite{BMNV} in two respects: i) we now consider also the contribution coming from the aberration effect at the observer's position, and ii)  we correct some minor errors (present in $d_L$ only at second order) due to the slightly different ``integration constants"
considered in \cite{BMNV} (see comments in Sect. 5).
However, we underline that such improvements affect in a totally negligible way the backreaction effects calculated in \cite{BGMNV2,BGMNV3}. 

Writing the result in the following concise form 
\beq
\frac{d_L(z_s, \tilde{\theta}^a)}{(1+z_s)a_o \Delta \eta}
= {d_L(z_s, \tilde{\theta}^a)\over d_L^{FLRW}(z_s)} =  1 + \bar{\delta}_S^{(1)}(z_s, \tilde{\theta}^a) + \bar{\delta}_S^{(2)}(z_s, \tilde{\theta}^a) ~~,
\eeq
 the first order luminosity distance  is given by
\be
\bar{\delta}_S^{(1)}(z_s, \tilde{\theta}^a) = \Xi_s \left(\partial_+ Q_s-\partial_r P_s\right)-\frac{1}{\Hcal_s\Delta \eta} \partial_r P_o - \frac{Q_s}{\Delta \eta} - \psi_s^{(1)} - J_2^{(1)} ~~. 
\label{finaldLord1}
\ee
We then have the following straightforward physical interpretation of the above terms: $\psi_s$ is a ``boundary term", while
\be 
-\frac{Q_s}{\Delta_\eta}=2 \int_{\eta^{(0)}_s}^{\eta_o} d\eta' \psi(\eta', \eta_o-\eta', \tilde{\theta}^a)
\ee
\be 
\partial_+ Q_s = \psi_o-\psi_s-2 \int_{\eta^{(0)}_s}^{\eta_o} d\eta' \partial_{\eta'}\psi(\eta', \eta_o-\eta', \tilde{\theta}^a)
\ee
are Sachs-Wolfe (SW) and integrated Sachs-Wolfe (ISW) effects, 
\be 
\partial_r P = \int_{\eta_{in}}^\eta d\eta' \frac{a(\eta')}{a(\eta)} \partial_r \psi(\eta',r,\tilde{\theta}^a) = \vec{v} \cdot \hat{n}
\ee
are Doppler effects (see also Eq.(\ref{GenVel})), and
\be 
 J^{(1)}_2 = \frac{1}{2}\left[\cot \theta ~ \tilde{\theta}^{(1)}+\partial_a \tilde{\theta}^{a (1)}\right] = 
 \frac12 \nabla_a \tilde{\theta}^{a (1)} = \frac{1}{\Delta\eta} \int_{\eta_s^{(0)}}^{\eta_o} d \eta' \,\frac {\eta' - \eta_s^{(0)}}{\eta_o - \eta'} \Delta_2 \psi(\eta', \eta_o-\eta', \tilde{\theta}^a) \,,
 \ee
is the first order lensing effect.

Following the pioneering work of \cite{Sasaki}, $d_L$ has been already computed to first order in the longitudinal gauge,
for a CDM model in \cite{Bonvin:2005ps} and for CDM and $\Lambda$CDM in \cite{Pyne:2003bn}. 
In particular,  we have verified that the first order result in Eq.(\ref{finaldLord1}), for the case of a CDM-dominated Universe,
is in full agreement with the result of \cite{Bonvin:2005ps}.

To second order we have a much more involved result.
For example, several terms arise from the fact that some of the first order terms in Eq.(\ref{finaldLord1}) have now to be integrated along the perturbed line of sight (see \cite{BMNV}). This gives rise to new terms which are given by the old ones multiplied by Doppler, SW and ISW effects.
As in \cite{BMNV} we choose to split the second order result in three different parts: 
\be
\bar{\delta}_S^{(2)}(z_s, \tilde{\theta}^a) = \bar{\delta}_{path}^{(2)} +  \bar{\delta}_{pos}^{(2)} +  \bar{\delta}_{mixed}^{(2)} \,,
\label{finaldLord2} 
\ee
where $\bar{\delta}_{path}^{(2)}$ denotes terms connected to the photon path and captures all the second order result in the absence of peculiar velocity effects; $\bar{\delta}_{pos}^{(2)}$ is for the terms generated by the source and observer peculiar velocity and captures all the  second order pure Doppler effects. Finally, $\bar{\delta}_{mixed}^{(2)}$ mixes peculiar velocity effects with all others. Their explicit expressions are:
\bea
\bar{\delta}_{path}^{(2)} &=& \Xi_s \Bigg\{ - \frac{1}{4} \left( \phi_s^{(2)} - \phi_o^{(2)} \right) + \frac{1}{4} \left( \psi_s^{(2)} - \psi_o^{(2)} \right) + \frac{1}{2} \psi_s^2 -  \frac{1}{2} \psi_o^2
- (\psi_s + J_2^{(1)} ) \partial_+ Q_s  
 \nonumber \\
&+& \frac14 (\gamma_{0}^{ab})_s \partial_a Q_s  \partial_b Q_s + Q_s \left( - \partial_+^2 Q_s + \partial_+ \psi_s \right)  
+ \frac{1}{{\mathcal H}_s} \partial_+ Q_s  
\, \partial_\eta \psi_s   \nonumber \\
&+& \frac14 \int_{\eta_o}^{\eta_s^{(0)-}} dx~ \partial_+ \left[ {\phi}^{(2)} + {\psi}^{(2)} + 4 {\psi} ~ \partial_+ Q + {\gamma}_{0}^{ab} ~ \partial_a Q ~ \partial_b Q \right] (\eta_s^{(0)+},x,\tilde{\theta}^a) \nonumber \\
&-& \frac{1}{2} \partial_a (\partial_+ Q_s) \, \left( \int_{\eta_o}^{\eta_s^{(0)-}} dx ~ \left[ {\gamma}_0^{ab} ~ \partial_b Q \right] (\eta_s^{(0)+},x,\tilde{\theta}^a) \right)
\Bigg\} \nonumber \\
&-& \frac{1}{2}\psi_s^{(2)} - \frac{1}{2} \psi_s^2 - K_2 + \psi_s J_2^{(1)} +\frac{1}{2}(J_2^{(1)})^2 
+ J_2^{(1)} \frac{Q_s}{\Delta \eta}-\frac{1}{\Hcal_s \Delta\eta} \left( 1 - \frac{\Hcal_s'}{\Hcal_s^2} \right) \frac12 (\partial_+ Q_s )^2 
\nonumber \\
&-& \frac{2}{\Hcal_s \Delta \eta} \psi_s \partial_+ Q_s + \frac12 \partial_a \left( \psi_s + J_2^{(1)} + \frac{Q_s}{\Delta \eta} \right)   \left( \int_{\eta_o}^{\eta_s^{(0)-}} dx ~ \left[ {\gamma}_0^{ab} ~ \partial_b Q \right] (\eta_s^{(0)+},x,\tilde{\theta}^a) \right) 
\nonumber
\\
&+& \frac14 \partial_a Q_s  \partial_+ \left( \int_{\eta_o}^{\eta_s^{(0)-}} dx ~ \left[ {\gamma}_0^{ab} ~ \partial_b Q \right] (\eta_s^{(0)+},x,\tilde{\theta}^a) \right)
\nonumber \\
&+& \frac{1}{16} \partial_a \left( \int_{\eta_o}^{\eta_s^{(0)-}} dx ~ \left[ {\gamma}_0^{bc} ~ \partial_c Q \right] (\eta_s^{(0)+},x,\tilde{\theta}^a) \right) \partial_b \left( \int_{\eta_o}^{\eta_s^{(0)-}} d\bar{x} ~ \left[ {\gamma}_0^{ad} ~ \partial_d Q \right] (\eta_s^{(0)+},\bar{x},\tilde{\theta}^a) \right) \nonumber \\
&-& \frac{1}{4 \Delta \eta} \int_{\eta_o}^{\eta_s^{(0)-}} dx~ \left[ {\phi}^{(2)} + {\psi}^{(2)} + 4 {\psi} ~ \partial_+ Q + 
{\gamma}_{0}^{ab} ~ \partial_a Q ~ \partial_b Q \right] (\eta_s^{(0)+},x,\tilde{\theta}^a) \nonumber \\
&+& \frac{1}{\Hcal_s} \partial_+ Q_s \left\{ - \partial_\eta \psi_s + \partial_r \psi_s + \frac{1}{\Delta \eta^2} \int_{\eta_s^{(0)}}^{\eta_o} d\eta' \Delta_2 \psi (\eta', \eta_o - \eta', \tilde{\theta}^a) \right\} \nonumber \\
&+& Q_s \left\{ \partial_r \psi_s + \partial_+ \left(\int_{\eta_o}^{\eta_s^{(0)-}} d x \frac{1}{(\eta_s^{(0)+}-x)^2} \int_{\eta_o}^x d y \Delta_2 {\psi}(\eta_s^{(0)+}, y, \tilde{\theta}^a) \right)\right. \nonumber \\
&+& \left. \frac{1}{2 \Delta\eta^2} \int_{\eta_s^{(0)}}^{\eta_o} d \eta' \Delta_2 \psi(\eta', \eta_o-\eta', \tilde{\theta}^a)
 \right\} \nonumber \\
&+& 
\frac{1}{16 \sin^2 \tilde{\theta}} \left( \int_{\eta_o}^{\eta_s^{(0)-}} dx ~ \left[{\gamma}_0^{1b} ~ \partial_b Q \right] (\eta_s^{(0)+},x,\tilde{\theta}^a) \right)^2 \, ,
\eea
\bea
\bar{\delta}_{pos}^{(2)} &=& \frac{\Xi_s}{2} \Bigg\{ \left(\partial_r P_s\right)^2 + (\gamma_0^{ab})_s \partial_a P_s \, \partial_b P_s
- \frac{2}{\Hcal_s} \left( \partial_r P_s - \partial_r P_o \right) \left( \Hcal_s \partial_r P_s + \partial_r^2 P_s \right) \nonumber \\
 &-& \int_{\eta_{in}}^{\eta_s^{(0)}} d\eta' \frac{a(\eta')}{a(\eta_s^{(0)})} \partial_r \left[ \phi^{(2)} - \psi^2 + (\partial_r P)^2 + \gamma_0^{ab} \partial_a P  \partial_b P \right] (\eta',\Delta\eta,\tilde{\theta}^a) 
\Bigg\} \nonumber \\
&+& \frac{1}{2 \Hcal_s \Delta \eta} \left\{ \left(\partial_r P_o\right)^2 +\lim_{r\rightarrow 0} \left[\gamma_0^{ab} \partial_a P \partial_b P \right]
\right. \nonumber \\
&-& \left. \int_{\eta_{in}}^{\eta_o} d\eta' \frac{a(\eta')}{a(\eta_o)} \partial_r \left[ \phi^{(2)} - \psi^2 + (\partial_r P)^2 + \gamma_0^{ab} \partial_a P \partial_b P \right] (\eta',0,\tilde{\theta}^a)\right\}
\nonumber 
\\
&-& \frac{1}{2 \Hcal_s \Delta\eta} \left( 1 - \frac{\Hcal_s'}{\Hcal_s^2} \right) \left( \partial_r P_s - \partial_r P_o \right)^2 \, ,
\eea
\bea
\bar{\delta}_{mixed}^{(2)} &=& \Xi_s \Bigg\{\partial_r P_s  J_2^{(1)}-\left(
\partial_r P_s-\partial_r P_o \right) \frac{1}{\Hcal_s} \partial_\eta \psi_s - (\gamma_0^{ab})_s \partial_a Q_s \partial_b P_s \nonumber \\
&+& \frac{1}{\Hcal_s} \partial_+ Q_s \partial_r^2 P_s + Q_s  \partial_r^2 P_s  \nonumber \\
&+&
\frac{1}{2} \partial_a ( \partial_r P_s - \partial_r P_o ) \, \left( \int_{\eta_o}^{\eta_s^{(0)-}} dx ~ \left[ {\gamma}_0^{ab} ~ \partial_b Q \right] (\eta_s^{(0)+},x,\tilde{\theta}^a) \right)
\Bigg\}
\nonumber \\
&-& \frac{1}{\Hcal_s \Delta\eta}  \left(\psi_o - \psi_s - J_2^{(1)} \right) \partial_r P_o +\frac{Q_s}{\Delta \eta} \partial_r P_s
\nonumber \\
&+& \frac{1}{\Delta \eta} ( \partial_r P_s - \partial_r P_o ) \Bigg\{ \frac{1}{\Hcal_s} \left( 1 - \frac{\Hcal_s'}{\Hcal_s^2} \right) 
\partial_+ Q_s + \frac{2}{\Hcal_s} \psi_s\Bigg\} \nonumber \\
&+& \frac{1}{\Hcal_s} ( \partial_r P_s - \partial_r P_o ) \!\left\{ \partial_\eta \psi_s - \partial_r \psi_s - \frac{1}{\Delta \eta^2}\! \int_{\eta_s^{(0)}}^{\eta_o} d\eta' \Delta_2 \psi (\eta', \eta_o - \eta', \tilde{\theta}^a) \right\}\!.
\eea

The physical interpretation of the terms above is more tricky with respect to the first order case.
Let us give here only two simple examples (but see also \cite{BMNV}),
\be 
K_2 = \frac{1}{2}\left[\cot \theta ~ \tilde{\theta}^{(2)}+\partial_a \tilde{\theta}^{a (2)}\right] = \frac12 \nabla_a \tilde{\theta}^{a (2)}
\ee
is the pure second order lensing effect, while, from Eq.(\ref{GenVel}), we have that
\be 
-\partial_i \tau^{(2)} n^i = \int_{\eta_{in}}^{\eta} d\eta' \frac{a(\eta')}{a(\eta_s^{(0)})} \partial_r \left[ \phi^{(2)} - \psi^2 + (\partial_r P)^2 + \gamma_0^{ab} \partial_a P \partial_b P \right] (\eta',r,\tilde{\theta}^a) 
\ee
is the second order Doppler effect coming from the second order peculiar velocity (at the observer or at the source). 
More about the physical interpretation of the second order contribution to $d_L$ can be also found in \cite{Umeh:2012pn}, where a summary of another second order calculation of $d_L$ in the PG, but only for the particular case of $\Lambda$CDM model, is presented.

\end{appendix}

%%%%%%%%%%%%%%%%%%%%%%%%%%%%%%%%%%%%%%%%%%%%%%%%%%%%%%%%%%%%%%%%%%%%%%%%%%%%%%%%%%%%%%%%%%%%%%%%%%%%%%%%
%%%%%%%%%%%%%%%%%%%%%%%%%%%%%%%%%%%%%%%%%%%%%%%%%%%%%%%%%%%%%%%%%%%%%%%%%%%%%%%%%%%%%%%%%%%%%%%%%%%%%%%%%

\end{document}